\documentclass[aip,amsmath,amssymb,reprint]{revtex4-1}

\usepackage{amsbsy}
\usepackage[dvipsnames]{xcolor}

\usepackage{dcolumn}
\usepackage{bm}

\usepackage[utf8]{inputenc}
\usepackage[T1]{fontenc}
\usepackage{mathptmx}
\usepackage{graphicx}
\usepackage{bbm}

\newcommand{\erf}{\mathrm{erf}}

\begin{document}

\preprint{AIP/123-QED}





\title[Fractional noise in nanopores]{Intrinsic fractional noise in nanopores: The effect of reservoirs}

\author{S. Marbach}
\email{sophie@marbach.fr}
\affiliation{ 
Courant Institute for Mathematical Sciences, New York University, New York, NY, U.S.A.
}%
 \affiliation{CNRS, Sorbonne Universit\'{e}, Physicochimie des Electrolytes et Nanosyst\`{e}mes Interfaciaux, F-75005 Paris, France.}





\date{\today}

\begin{abstract}
Fluctuations affect nanoporous transport in complex and intricate ways, making optimization of signal-to-noise in artificial designs challenging. Here we focus on the simplest nanopore system, where non-interacting particles diffuse through a pore separating reservoirs. We find that the concentration difference between both sides (akin to the osmotic pressure drop) exhibits fractional noise in time $t$ with mean square average that grows as $t^{1/2}$. This originates from the diffusive exchange of particles from one region to another. We fully rationalize this effect, with particle simulations and analytic solutions. We further infer the parameters (pore radius, pore thickness) that control this exotic behavior. As a consequence, we show that the number of particles \textit{within} the pore also exhibits fractional noise. Such fractional noise is responsible for noise spectral density scaling as $1/f^{3/2}$ with frequency $f$, and we quantify its amplitude. Our theoretical approach is applicable to more complex nanoporous systems (for example with adsorption within the pore) and drastically simplifies both particle simulations and analytic calculus.
\end{abstract}



\maketitle 

\section{Introduction}

\subsection{General introduction}

Fluctuations are ubiquitous in biological and artificial nanopores. The nanopore structure~\cite{bezrukov2000examining,siwy2002origin,marbach2018transport}, its position on the membrane~\cite{lawley2019diffusive}, its inner physical properties such as the local surface charge~\cite{scalfi2020charge}, and finally the number of particles inside and outside of the pore are all inherent sources of fluctuations. Their consequences on nanoporous transport are intricate and leave in particular strong signatures in noise measurements of currents. For example, the fluctuations of ionic current through a nanopore usually exhibit strong frequency dependence at low frequencies. Typically, the power spectral density of the current scales as $S(f) \sim 1/f^{\alpha}$ where $\alpha = 0.5 - 2.0$ according to the specifics of the system. Such a power law dependence, generally referred to as \textit{low frequency noise}, has been measured repeatedly in biological pores~\cite{wohnsland19971,bezrukov2000examining,nestorovich2003residue,siwy2002origin} and in a great diversity of artificial nanopores.~\cite{siwy2002origin,dekker2007solid,smeets2008noise,smeets2009low,powell2009nonequilibrium,hoogerheide2009probing,tasserit2010pink,powell2011noise,heerema20151,secchi2016scaling,wen2017generalized,fragasso20191,knowles2019noise}

Understanding precisely the origin and magnitude of such noise is important for two reasons. First, to shed light on the transport mechanism and allow us to track single molecule events.~\cite{zevenbergen2007mesoscopic,zevenbergen2011stochastic,krause2014brownian} Second, to provide guidelines to optimize the signal-to-noise ratio to improve sensitivity of single molecule detection experiments or DNA sequencing.~\cite{clarke2009continuous,howorka2009nanopore,kowalczyk2010detection,bell2015specific} Most efforts on improving signal-to-noise ratio are experimental and have been directed towards developing multilayered~\cite{chen2004atomic,tabard2007noise,beamish2012precise}, surface treated pores to improve insulation~\cite{balan2014improving,chang2004dna} or adsorption effects.~\cite{knowles2020investigating} Yet theoretical advances are necessary to open new optimization avenues and improve our general understanding of noise in nanoporous transport.  


First principles theories for fluctuations in nanoscale systems have remained sparse as theoretical treatments face several challenges, such as solving equations in complex geometries~\cite{bezrukov2000particle,zorkot2018current,gravelle2019adsorption}, or  accounting for all the various interactions between solute particles.~\cite{zorkot2016current,zorkot2016power} Furthermore, noise on ionic currents does not result from a single effect but from a combination of various effects that are more or less important according to the system investigated.~\cite{powell2011noise}  Recent modeling advances have nonetheless pointed to the crucial role of adsorption inside the pores~\cite{gravelle2019adsorption,knowles2020investigating}, of ion-ion correlations~\cite{zorkot2016power}, and of ionic concentration.~\cite{zorkot2016current} Most importantly, quantification of the amplitude of different noise sources is seldom available. 

\begin{figure}[h!]
\includegraphics[width = 0.99\columnwidth]{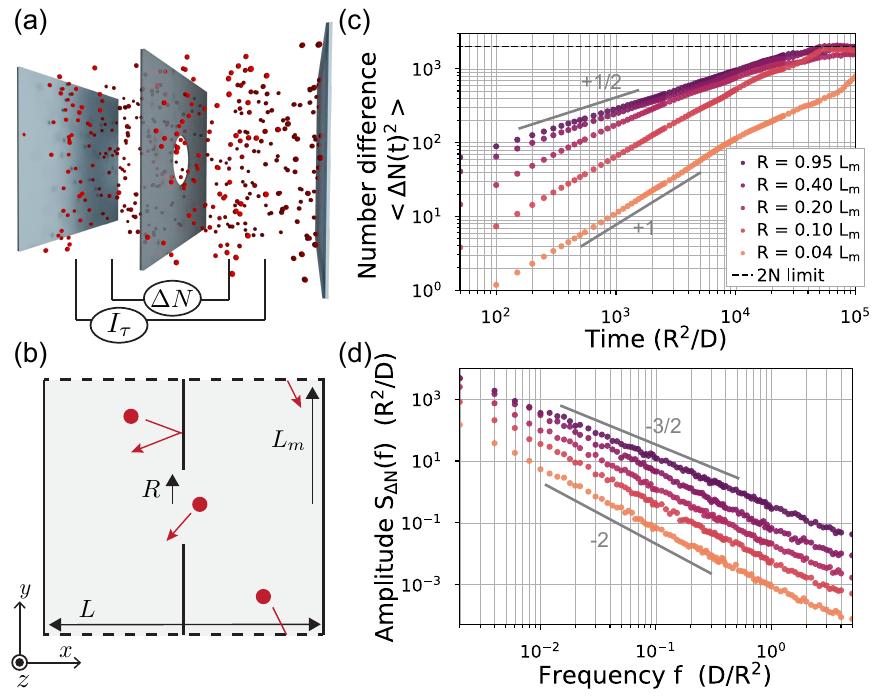}
\caption{\label{fig:fig1} \textbf{Fractional noise in a simple nanoporous system}. (a) Brownian dynamics of non-interacting particles (red spheres). Particles cross the wall through a circular pore of radius $R$. (b) Corresponding schematic with simulation parameters and details. Dashed (resp. full) lines denote periodic (resp. reflecting) boundaries. (c) Mean square concentration difference $\langle \Delta N^2(t) \rangle$ with time for different values of the pore size. (d) Corresponding frequency spectrum -- shared legend with (d). Data points correspond to $N = 1000$ brownian walkers simulated in a box of size $L = 500~R$. Values of $L_m$ with respect to $R$ are indicated in the labels. The total simulation time was $1.5 \times 10^{6} \, R^2/D$ with a time step $\Delta t = 0.05 \, R^2/D$.}
\end{figure}

Here we come back to basics and explore the simplest possible setting for nanoporous transport, with a focus on reservoir effects. We study non-interacting solute particles diffusing between two compartments separated by a membrane with a single pore -- see Fig.~\ref{fig:fig1}-a and~ b. We investigate relevant observables in this context: \textit{(i)} the concentration difference $\Delta N$ between the two compartments (akin to the osmotic pressure drop at small concentration differences~\cite{marbach2019osmosis}), \textit{(ii)} the current of (uncharged) particles crossing the membrane, and \textit{(iii)} the number of particles within the pore. We find that such a simple system features very non-trivial noise characteristics. For example, fluctuations of the concentration difference grow as a power law $\langle \Delta N^2(t) \rangle \sim t^p$ (and plateau at long times). The power law factor $p = 0.5 - 1$ according to the radius of the pore -- see Fig.~\ref{fig:fig1}-c. Notably, this results in a noise spectra $S(\Delta N, f) \sim 1/f^{1+p}$ -- see Fig.~\ref{fig:fig1}-d. We will show that similar power law dependencies are seen in the other observables (\textit{ii}-\textit{iii}). Importantly, such fractional noise (with $p = 0.5$) is reminiscent of an intrinsic mathematical property of Brownian walkers exchanging between two sides of an imaginary boundary on a line.~\cite{harris1965diffusion,durr1985asymptotics} Its consequences in the context of nanoporous transport have yet to be observed and rationalized. Interestingly, such power law dependencies have been repeatedly observed in experimental or theoretical observations albeit rarely explained.~\cite{bezrukov2000examining,siwy2002origin,wen2017generalized,gravelle2019adsorption,knowles2020investigating}






In this paper, we fully rationalize, theoretically and numerically, the emergence of fractional noise in these observables \textit{(i -- iii)}. Our numerical model is based on Brownian dynamics of non-interacting, uncharged, particles. Our analytic treatment relies on a mapping of the full 3D problem to a simpler 1D problem, preserving equilibrium properties. This allows to bypass geometric complexities~\cite{gravelle2019adsorption,zorkot2018current} and obtain analytic expressions. We show that the mapping solutions reproduce exactly Brownian dynamics simulations in 3D. It further builds a general numerical framework to account efficiently for the effect of reservoirs without the introduction of artificial pore entry rates.~\cite{zevenbergen2009electrochemical} Our analytic results shed light on the mechanisms at play. Interestingly, we find that $1/f^{3/2}$ noise spectra are predominantly seen in thick pores (akin to channels) while $1/f^{2}$ spectra correspond to thin pores. We further find a $1/f^{1/2}$ low frequency decay for the number of particles within the pore. Importantly, we are able to quantify their amplitude in terms of the parameters of the system (pore size, pore thickness). From these results we deduce rules to optimize signal-to-noise ratio in several cases. In particular, for currents associated with the number of particles within the pore (akin to number of charge carriers for charged species) we find that signal-to-noise is maximized for \textit{short} pores (in contrast with long pores).

\subsection{Setup to probe the effect of reservoirs}

In this study, we consider a simple nanoporous system, where particles -- representing the solute species -- may diffuse freely across a membrane pore of characteristic width $R$ set on a membrane square of size $2 L_m \times 2 L_m$ -- see Fig.~\ref{fig:fig1}-a and b. In Fig.~\ref{fig:fig1}-a the pore is a circular pore of radius $R$. The radius $R$ corresponds to the accessible pore radius. Our derivation is not limited to circular pores and can be easily extended to other cross-sections such as slits. Let $x$ be the direction orthogonal to the membrane plane and $y$ and $z$ along it. $x = y = z=0$ corresponds to the position of the pore center on the membrane. We consider periodic boundary conditions in $y$ and $z$ at distance $L_m$ from the pore center (dashed lines in Fig.~\ref{fig:fig1}-b). This means that the open area with respect to the total area of the membrane is $\pi R^2/4L_m^2$. When $R \sim L_m$, this corresponds to a large pore (or, making use of the periodic boundary conditions, a membrane with multiple nearby pores). When $R \ll L_m$ this corresponds to a small pore (or a membrane with isolated pores). 

The finite extent of the reservoirs is modeled by reflection boundary condition at $x = \pm L/2$ parallel to the membrane. Our simulation setting therefore allows to probe the effect of pore size and reservoir size on translocation processes. Unlike other studies introducing effective boundary conditions to model the effect of reservoirs~\cite{bezrukov2000particle,zevenbergen2009electrochemical}, here we directly probe the effect of the \textit{presence} of reservoirs on the system by fully accounting for them. Note that in the following, the most striking effects will arise from the presence of reservoirs, in that they allow exchanges of particles between pore and reservoir regions, and not as much from their finite extent.

The $N$ particles are modeled as Brownian walkers. The displacement of each walker during a time $\Delta t$ is given by
\begin{equation}
\Delta \bm{X}_{k} = \sqrt{2 D \Delta  t} \,\bm{W}_{k}
\label{eq:starting}
\end{equation}
where the $\bm{W}_{k}$ are Gaussian random variables with mean $\langle \bm{W}_{k} \rangle = 0$ and variance $\langle W^i_{k} W^j_{l} \rangle = \delta_{i,j} \delta_{k,l}$ and  $\langle . \rangle$ are averages over realizations of the noise. Here $\bm{X}_{k} = (x_k,y_k,z_k)$ is the position a walker where $k$ is the running index over time, such that time is $t = k \Delta t$. 

We introduce $\Delta N (t) = N_R (t) - N_L (t)$ as the difference between the number of particles to the right of the membrane (particles for which $x_k > 0$) versus particles to the left ($x_k < 0$). We are interested in the statistics of this random variable $\Delta N$, notably because it represents the concentration difference between both sides, and thus is linearly related to the osmotic pressure (at small concentration differences).~\cite{marbach2019osmosis}

\subsection{Summary}

The paper is organized as follows. 

In section~\ref{sec:fractionalOrigin}, we explain in details the emergence of fractional noise ($\langle \Delta N^2(t) \rangle \sim \sqrt{t}$) in the simple setting of a fully open membrane (corresponding to the large pore regimes, $R \sim L_m$, purple in Fig.~\ref{fig:fig1}-c and d). This setup is equivalent to studying brownian walkers on a line. We rationalize fluctuations, correlation functions and spectrum properties of both the number particle difference and current observables. The most important finding here is that fractional noise emerges spontaneously when studying \textit{random particles crossing from one region to another} -- here from the left to the right. As a consequence we expect fractional noise to be universal and emerge in many different settings, that we investigate in the following sections. 

In section~\ref{sec:narrowPores} we investigate how these results are maintained when the particles can only cross through narrow pores. We also introduce the mapping of the 3D geometry to a 1D problem. This allows us to fully rationalize the different behaviors obtained in Fig.~\ref{fig:fig1}-c and~d. The key takeaway here is that different behaviors are obtained not only with the pore size but especially in time. In general, in contrast with large pores, small pores $R \ll L_m$ exhibit diffusive noise over longer time scales. Yet within specific time windows \textit{fractional noise may also be observed across all pore sizes}. 

In section~\ref{sec:nanochannels} we investigate how fractional noise impacts long channels. Importantly, when the pore is long, it is possible to track the number of particles within the pore, akin to the number of charge carriers responsible for ionic conductance in ionic systems. We focus mainly on the noise properties of this number in this final section. Because of its intrinsic nature, fractional noise is also observed in the number of particles within the pore. Interestingly, the noise spectrum of the number of particles within the pore exhibits not only power laws expected for fractional noise but an additional $1/f^{1/2}$ power law scaling over a range of smaller frequencies. 

In all sections we discuss the results with the aim of optimizing signal-to-noise ratio. 

\section{Origin of fractional noise}
\label{sec:fractionalOrigin}

\subsection{Limit case of a wide pore: walkers on a line}

To understand the emergence of fractional noise in nanopores, we focus first on a limit case. Fig.~\ref{fig:fig1}-c shows that fluctuations in the number difference grow as $\langle \Delta N^2(t) \rangle \sim t^{1/2}$ predominantly for wide pores $R \sim L_m$. 
In this limit ($R \sim L_m$) we can consider as a first approximation that there is no physical membrane. Solute particles are diffusing and we consider their probability of crossing the now "imaginary" wall at $x =0$. Everything now happens as if the particles where walking on a line -- see Fig.~\ref{fig:singleFile}-a. 

\subsection{Number difference}

In average  $\langle \Delta N (t) \rangle = 0$. To quantify further the fluctuations of $\Delta N (t)$ we therefore turn to its correlation function $\langle \Delta N(t) \Delta N(0) \rangle$. 

$\Delta N $ can be conveniently written as $\Delta N (t+\Delta t) = \Delta N (t) + 2 I_{\Delta t} (t) \Delta t$ where $I_{\Delta t}(t)$  corresponds to the \textit{net} current of (uncharged) particles crossing the boundary $x=0$ between $t$ and $t+\Delta t$. A particle that started in $x<0$ (resp. $x>0$) at time $t$ and finds itself in $x>0$ (resp. $x<0$) at time $t+\Delta t$ will contribute $+1/\Delta t$ (resp. $-1/\Delta t$) to the current.
Note that the definition of such a current does not pose any mathematical pathology. Although a Brownian walker does cross a boundary infinitely many times during $\Delta t$, here $I_{\Delta t}(t)$ is finite since it counts whether the walker effectively crossed (for example a particle crossing 3 times back and forth will contribute $+1-1+1 = 1$ time to the current). 

Summing up over time we obtain $\displaystyle \Delta N (t) = \Delta N (0) + 2 \sum_{k'=0 \ldots k} I_{\Delta t} (k'\Delta t) \Delta t \rightarrow \Delta N(0) + 2 \int_0^t I(t_1) dt_1$ in the limit of small time steps.  We can therefore write the correlation function for $\Delta N$ as 
\begin{equation}
\begin{split}
C(t,t') &= \langle (\Delta N(t) - \Delta N(0)) (\Delta N(t') - \Delta N(0)) \rangle \\
&=  4 \langle \int_0^t I(t_1) dt_1 \int_0^{t'} I(t_2) dt_2 \rangle.
\end{split}
\end{equation}
We will write $\langle \Delta N^2(t) \rangle = \langle (\Delta N(t) - \Delta N(0))^2 \rangle$ in the following to lighten notations. 

\subsection{Statistics of the number difference}

To determine the statistics of $\Delta N(t)$ it is therefore sufficient to find the statistics of $I(t)$. In the following, we will use a number of standard results for diffusing tracers in one dimension (for detailed proofs of these results, see Chapters 2 and 3 of Ref.~\onlinecite{crank1979mathematics}). We compute statistics by splitting the calculation in two parts:
\paragraph*{Jumps on a common interval.} We are interested first in the correlation function at equal times
\begin{equation}
         C_{\rm common}(t) = 4 \langle \int_0^{t} I(t_1) dt_1\int_0^{t} I(t_2) dt_2 \rangle. 
         \label{eq:Common}
\end{equation}
As particles are uncorrelated we can focus on a single particle.    

We first derive the probability for the current to be $+1/\Delta t$ during $\Delta t$; meaning that the particle crossed from left to right. At any time $t$ the particle is evenly distributed between left and right side with a distribution $\rho_0(x) = 1/L$, where $L$ is the domain size. In the following we will assume $t \ll L^2/D$ to neglect boundary effects due to the finite extent of reservoirs. The full derivation accounting for those effects is reported in Appendix B and shows no difference at these timescales. 

The probability that a step has size $\Delta x$ during $\Delta t$ is 
\begin{equation}
    p(s = \Delta x) = \frac{1}{\sqrt{4\pi D \Delta t}} e^{- \frac{\Delta x^2}{4 D \Delta t}}
    \label{eq:probaSteps}
\end{equation}
and therefore the probability that the particle made a step greater than $\Delta x $ is 
\begin{equation}
    p(s \geq \Delta x) = \int_{ \Delta x}^{\infty} p(s = s') ds' = \frac{1}{2} \left( 1 - \mathrm{erf} \left(  \frac{\Delta x}{\sqrt{4 D \Delta t}}\right) \right)
\end{equation}
where here the upper integration bound is not $L$ but $+\infty$ as we neglect the finite extent of reservoirs. The probability that the current is +1 is equal to the probability that a particle came from the left \textit{and} made it to the right  
\begin{equation}
p\left(I_{\Delta t} = +1/\Delta t\right) =   \int_{-\infty}^{0} \rho_0(x) p(s \geq -x) dx = \frac{1}{L} \sqrt{\frac{D \Delta t}{\pi}}.
\end{equation}
Since the probability to observe current in one direction or the other is the same we have $p(I_{\Delta t} = -1/\Delta t) = p(I_{\Delta t} = +1/\Delta t) $. In average the current vanishes naturally, $  \langle I (t) \rangle = 0$. 

\begin{figure}[h!]
\includegraphics[width = 0.99\columnwidth]{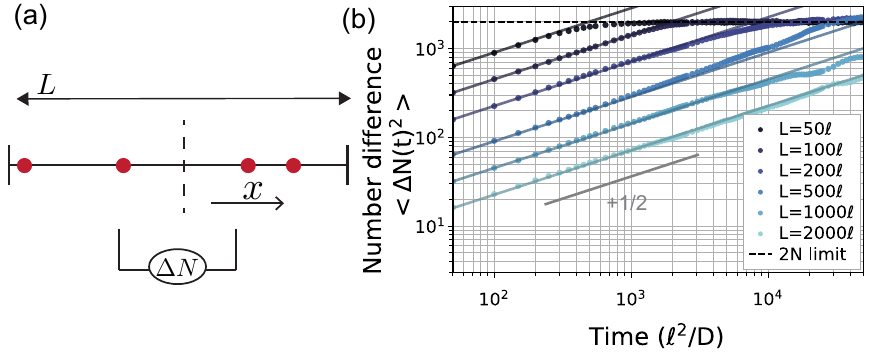}
\caption{\label{fig:singleFile} \textbf{Fractional noise for the concentration difference on a line}. (a) Illustration of random walkers on a line. Red particles can cross the imaginary boundary at the center. (b) Mean square concentration difference $\langle \Delta N^2(t) \rangle$ with time for different values of box size $L$. Dots correspond to BD simulations and full lines to Eq.~\eqref{eq:growthWitht12}. Here $N = 1000$ particles were simulated and $\ell$ is an arbitrary unit length. The total simulation time was $1.5 \times 10^{6} \, \ell^2/D$ with a time step $\Delta t = 0.05 \, \ell^2/D$.}
\end{figure}

If now we consider a longer time interval $t$, the derivation does not change, and we can simply replace $\Delta t \rightarrow t$. The probability that the current integrated over time $t$ is $+1$ is thus
\begin{align}
   p\left(\int_0^t I(t_1) dt_1 = +1\right)  =   \frac{1}{L} \sqrt{\frac{D t}{\pi}}   
   \label{eq:pLeft}
\end{align}
and similarly for the reverse current. The equal time correlation for one particle therefore amounts to
\begin{equation}
\begin{split}
    \langle \int_0^{t} I(t_1) dt_1\int_0^{t} I(t_2) dt_2 \rangle =  & (-1)^2 p\left(\int_0^t I(t_1) dt_1 = -1\right) \\
    &+ (1)^2 p\left(\int_0^t I(t_1) dt_1 = +1\right).  \\
    \end{split}
\end{equation}
Using Eqs.~\eqref{eq:Common} and~\eqref{eq:pLeft} and multiplying by the number of (independent) particles, we find
\begin{equation}
         \langle \Delta N(t)^2 \rangle =   C_{\rm common}(t) =  8 \frac{N}{L} \sqrt{\frac{D}{\pi}} \sqrt{t}.
         \label{eq:growthWitht12}
\end{equation}
As expected, concentration fluctuations scale as $\langle \Delta N(t)^2 \rangle/N^2 \sim 1/N$. In small sized systems such as cells or nanofiltration devices we therefore expect these fluctuations to be significant. The growth law Eq.~\eqref{eq:growthWitht12} agrees perfectly with Brownian dynamics (BD) simulations, for very different numerical parameters -- see Fig.~\ref{fig:singleFile}-b and Appendix A for simulation details. Interestingly Eq.~\eqref{eq:growthWitht12} shows that concentration fluctuations have large deviations with time. These deviations are eventually bounded by the system size -- we turn to this limit next.

\vspace{2mm}

\paragraph*{Limit value of fluctuations.} At long enough times, the fluctuations are bounded. Indeed, at long times we can write 
$N_L = n$ and $N_R = N - n$ 
where $n$ is a Binomial random variable. A total of $N$ particles are placed on either side of the membrane with equal probability $p = 1/2$. $n$ has mean value $\langle n \rangle = p N = N/2$ and variance $\langle (n - \langle n \rangle)^2 \rangle = N p (1 - p) = N/4$. Thus we can calculate
\begin{equation}
\label{eq:limit2N}
\begin{split}
\langle \Delta N^2 \rangle &= \langle (N_R - N_L)^2 \rangle = \langle (N - 2n)^2 \rangle = 2N.
\end{split}
\end{equation}
The limit law Eq.~\eqref{eq:limit2N} is consistently seen in our simulations -- see Fig.~\ref{fig:singleFile}-b. 

The time $t_{\rm late} $ to reach saturation is set by equating Eqs.~\eqref{eq:growthWitht12} and~\eqref{eq:limit2N} such that 
$t_{\rm late} = \frac{\pi}{16} \frac{L^2}{D}$.
Naturally this time corresponds to the typical time $\frac{L^2}{D}$ to diffuse to the boundaries of the domain. In experimental systems, typically the size of the reservoirs is $L = 1~\mathrm{cm}$ and $D \sim 2\times 10^{-9} \mathrm{m^2/s}$ yielding $t_{\rm late} \sim 1~\mathrm{h}$. Such square root noise dependence is therefore accessible to experimental systems.

\vspace{2mm}

\paragraph*{Jumps on consecutive intervals.} We are now interested in jumps on consecutive intervals of the correlation function, as 
\begin{equation}
         C_{\rm consecutive}(t_1,t_2) = \langle \int_0^{t_1} I(t) dt\int_{t_1}^{t_1+t_2} I(t') dt' \rangle 
\end{equation}
where the first interval is of length $t_1$ and the second of length $t_2$. For simplicity we will write in this section $I^{(i)}$ the current on the $i^{\rm th}$ interval. We focus as before on one particle. As we have shown in the previous paragraph, the probability that the integrated current is $+1$ during a length $t_1$ is $ p\left(I^{(1)} = +1 \right) = \sqrt{D t_1/\pi}$. After that first jump the particle is distributed as
    \begin{equation}
        \rho_1(x) = \frac{1}{2L} \sqrt{\frac{\pi}{D t_1}} \left( 1 - \mathrm{erf}\left( \frac{x}{\sqrt{4Dt_1}} \right)\right)
    \end{equation}
such that the probability that the particle crosses again (in the reverse direction) during the second time lapse $t_2$ is 
\begin{equation}
    \begin{split}
        p\left( I^{(2)} = -1 \big|I^{(1)}  = +1 \right) &= \int_0^{\infty} \rho_1(x) p(s \leq -x, t_2) dx \\
        &= \frac{1}{2 L} \left( 1 + \sqrt{\frac{t_2}{t_1}} - \sqrt{1- \frac{t_2}{t_1}} \right).
\end{split}
\end{equation}
 The current correlations can be expressed in terms of the jump probabilities as
\begin{equation}
    \begin{split}
         \langle I^{(2)} I^{(1)}   \rangle  =&-  p(I^{(1)}  = +1)  p(I^{(2)} = -1 \big|I^{(1)}  = +1 ) \\
         &-  p(I^{(1)}  = -1)  p(I^{(2)} = +1 \big|I^{(1)}  = -1 ) 
         \end{split}
\end{equation}
and coming back to $N$ particles we find
\begin{equation}
    C_{\rm consecutive}(t_1,t_2) = - 4 \frac{N}{L} \sqrt{\frac{D}{\pi}} \left( \sqrt{t_1} + \sqrt{t_2}  - \sqrt{t_1 + t_2}\right).
\end{equation}

\paragraph*{Statistics of $\Delta N$.}

Assembling the different jumps we have 
$
C(t,t') = C_{\rm common}(t)  + C_{\rm consecutive}(t,t'-t)
$, writing without loss of generality $t' > t$.
We obtain 
\begin{equation}
    \langle \Delta N(t) \Delta N(t') \rangle = 4 \frac{N}{L}  \sqrt{\frac{D}{\pi}} \left( \sqrt{t} + \sqrt{t'} - \sqrt{t' - t}\right) .
    \label{eq:DeltaNcorr}
\end{equation}
Eq.~\eqref{eq:DeltaNcorr} is exactly the expectation value of a fractional Brownian walk~\cite{mandelbrot1968fractional} with Hurst index $H = 1/4$. This allows to conclude that $\Delta N$ is a fractional Brownian walk with "diffusion coefficient" $\mathcal{D} = 4 \frac{N}{L}  \sqrt{\frac{D}{\pi}}$. Notably, the emergence of such fractional Brownian noise is totally intrinsic and relies on no specific assumptions for the system. As such, it could serve as a remarkable textbook example for fractional or subdiffusive noise. 

Note that Eq.~\eqref{eq:DeltaNcorr} can be inferred in many different ways.~\cite{harris1965diffusion,durr1985asymptotics} The proof presented here -- in contrast with other more formal proofs -- sheds light on the \textit{physical mechanisms} that result in such peculiar statistics. \textit{Namely, particles crossing forwards and in a limited amount of time turning around and crossing back.} This is also at the basis of the current statistics, which we study in the following section.  

\subsection{Current of (uncharged) particles}

As most experimental apparatus are sensitive not to an instantaneous current but to a current integrated over a short time interval say $\tau$, we define the experimentally relevant current (of uncharged particles) as
\begin{equation}
I_{\tau}(t) = \int_t^{t+\tau} I(t_1)dt_1.
\end{equation}
Typicallly $\tau^{-1} \sim 100~\mathrm{kHz}$. In our non-dimensional time scales, with a typical length scale for nanopores $\ell \sim 10~\mathrm{nm}$ and $D = 10^{-9}~\mathrm{m^2/s}$ we have $\tau \sim 100 \frac{\ell^2}{D}$. 

\begin{figure}[h!]
\includegraphics[width = 0.99\columnwidth]{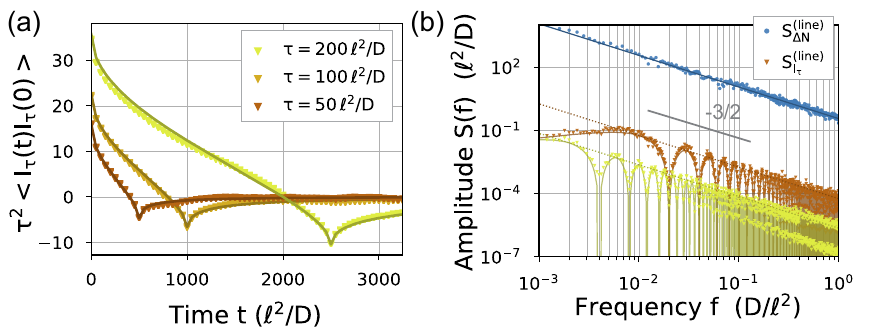}
\caption{\label{fig:singleFileFourier} \textbf{Current fluctuations (of uncharged particles) on a line}. (a) Autocorrelation function of the current from BD simulations on a line (dots, multiplied by the integration time $\tau^2$ for more clarity on a single plot). The full lines correspond to Eq.~\eqref{eq:CorrelationCountNoFlux}. (b) Spectrum of fluctuations for the number difference $\Delta N$ and two of the current traces $I_{\tau}$ shown in (a). The full blue line corresponds to Eq.~\eqref{eq:SdN} and the full maroon and green lines to Eq.~\eqref{eq:SiT} (the dashed lines correspond to Eq.~\eqref{eq:SiT} taking the sinusoidal multiplicative factor $ \sin^2(2\pi f \tau) \equiv 1$ to show the power law decay). Numerical parameters correspond to that of Fig.~\ref{fig:singleFile}.}
\end{figure}

We now seek the correlations of $I_{\tau}(t)$. When $t \leq \tau$ we can split the correlation function calculation as 
\begin{align}
\begin{split}
     \langle&  I_{\tau}(t) I_{\tau}(0) \rangle  = \frac{1}{\tau^2} \bigg(  \langle \int_t^{\tau} I(t_1) dt_1\int_t^{\tau} I(t_2) dt_2 \rangle + \\
     &  \langle \int_t^{\tau} I(t_1) dt_1\int_0^{t} I(t_2) dt_2 \rangle + \langle \int_{\tau}^{t + \tau} I(t_1) dt_1\int_0^{\tau} I(t_2) dt_2 \rangle \bigg)
     \end{split}
\end{align}
such that we can use previous results on correlation functions on common and consecutive intervals. 
When $t > \tau$ another splitting may be done leading to a similar result
such that the correlation is simply (for any time $t$)
\begin{equation}
   \langle  I_{\tau}(t) I_{\tau}(0) \rangle = \frac{N}{L \tau^2} \sqrt{\frac{D}{\pi}}  \left(\sqrt{|t - \tau|} + \sqrt{t + \tau} - 2 \sqrt{t}  \right).
    \label{eq:CorrelationCountNoFlux}
\end{equation}
Eq.~\eqref{eq:CorrelationCountNoFlux} corresponds to a result derived in Ref.~\onlinecite{burov2011single}, for the velocity autocorrelation function of a fractional Brownian walker -- up to subtleties associated with the different systems. Here, current $I(t)$ plays the role of the velocity, since it is the derivative of the number difference
$
\frac{d\Delta N(t)}{dt} = 2 I(t)
$.
Eq.~\eqref{eq:CorrelationCountNoFlux} agrees perfectly with BD simulations -- see Fig.~\ref{fig:singleFileFourier}-a. Notably, it features a \textit{negatively correlated} peak. This peak corresponds to particles that cross, turn around and cross back again. The maximum value of the negative peak is achieved in $t = \tau$  and is of significant magnitude since
$
\frac{\langle  I_{\tau}(t) I_{\tau}(0) \rangle}{\langle I_{\rm \tau}(0)^2 \rangle} =  \simeq - 0.3
$ regardless of the system specifics. 

We now turn to exploring the typical signature of the noise in the current spectrum.

\subsection{Spectrum of fluctuations}

We define the spectrum of a random variable $X$ as 
\begin{equation}
S_X(f) = \lim_{T\rightarrow \infty} \left \langle \frac{1}{T} \bigg| \int_0^{T} e^{i 2\pi f t} X(t) dt\bigg|^2 \right \rangle.
\end{equation}
Our simulations are long enough that we need not account for finite acquisition time effects.~\cite{krapf2019spectral}

\paragraph{Number difference spectrum.}

As the number difference $\Delta N$ may drive currents in more complex systems (\textit{e.g.} osmotic currents), we first investigate its spectrum. From Eq.~\eqref{eq:DeltaNcorr} we obtain:~\cite{krapf2019spectral}
\begin{equation}
\label{eq:SdN}
S^{\rm (line)}_{\Delta N}(f) = \frac{2 N \sqrt{2 D}}{L} \frac{1}{(2\pi)^{3/2}} \frac{1}{f^{3/2}}.
\end{equation}
The spectrum of the noise therefore has a low frequency decay as $1/f^{3/2}$ -- consistently seen in simulations, see Fig.~\ref{fig:singleFileFourier}-b (blue curves). 

\paragraph{Spectrum of particle current.} The spectrum of the particle current is easily calculated from Eq.~\eqref{eq:CorrelationCountNoFlux} as 
\begin{equation}
\label{eq:SiT}
S^{\rm (line)}_{I_{\rm tau}}(f) = \frac{N}{L \tau^2} \sqrt{\frac{D}{\pi}}  \frac{\sin^2(2\pi f \tau)}{(2\pi)^{3/2}} \frac{1}{f^{3/2}}.
\end{equation}
We find again a low frequency decay as $1/f^{3/2}$, also seen in simulations, see Fig.~\ref{fig:singleFileFourier}-b. 

\vspace{3mm}

Statistics of particles on a single line decisively point to a low frequency noise scaling as $1/f^{3/2}$. This noise simply originates from particles crossing from one region to another. In nanoporous systems, particles continuously cross from the reservoir to the pore area and back. We therefore expect such \textit{intrinsic} fractional noise to leave traces. Nanoporous systems are yet more complex than the simple line problem. In the following sections we investigate how these results hold or change in more realistic geometries.

\section{Fractional noise in narrow pores}
\label{sec:narrowPores}

When the pore mouth is extremely small, a regime different from fractional Brownian noise is expected. In fact, we expect the \textit{return probability} of crossing particles to vanish, as the opening is so narrow that particles can not find it again in finite time. Simulations show that $\langle \Delta N^2(t) \rangle \sim t^{0.5-1}$. However there is no systematic way of understanding the fluctuations of $\Delta N$ for narrow pores. This is the purpose of the following section. 

\subsection{Mapping to a simpler problem}

\begin{figure}[h!]
\includegraphics[width = 0.99\columnwidth]{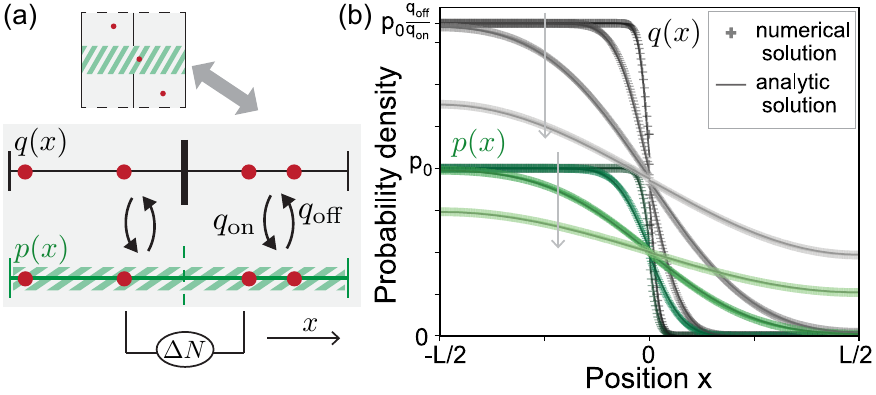}
\caption{\label{fig:mapping} \textbf{Mapping of a 3D problem to the -- easy to solve -- rates model}. (a) Sketch of the model: walkers (red) perform a Brownian walk on two adjacent lines. On the black line, walkers cannot cross at the center. Walkers can swap lines with rates $q_{\rm on}$ and $q_{\rm off}$. (b) Probability density for the particles on the crossing line ($p(x)$) and on the blocked line ($q(x)$) starting on the left. Finite difference numerical solutions (dots, see Appendix A) and analytic solutions of Eq.~\eqref{eq:ratesProblem} are presented. Arrows indicate the direction of time and profiles are represented respectively at $t = 50,\,500,\, 5000,\, 25000\, R^2/D$. Here $2 R = 3 L_m$ such that $q_{\rm off} \simeq 1.86 q_{\rm on}$. }
\end{figure}

We focus on the case of short pores -- as depicted in Fig.~\ref{fig:fig1}. Seeking solutions as in Sec.~\ref{sec:fractionalOrigin} of the full problem is tedious and greatly dependent on the specific geometry of the pore. Instead, we map the open pore problem to a simpler model that features similar equilibrium characteristics -- see Fig.~\ref{fig:mapping}-a. 

In this setup -- termed henceforth the \textit{rates model} -- a particle performs a random walk freely on a \textit{passing} line (corresponding to the open pore cylinder -- see shaded green Fig.~\ref{fig:mapping}-a). The particle can transition with rate $q_{\rm off}$ to a \textit{blocked}  line with a reflecting wall in the center (corresponding to the membrane wall). The particle transitions back to the open line with rate $q_{\rm on}$. The rates have to obey detailed balance
\begin{equation}
    q_{\rm on} \mathcal{A}_{\rm closed} = q_{\rm off} \mathcal{A}_{\rm open}
\end{equation}
where $\mathcal{A}_{\rm i}$ are the areas corresponding to the closed or open parts of the membrane. In the case of the circular pore of radius $R$ on a membrane square of size $2 L_m \times 2 L_m$ we get
\begin{equation}
    \frac{q_{\rm on}}{q_{\rm off}} = \frac{\mathcal{A}_{\rm open}}{\mathcal{A}_{\rm closed}} = \frac{\pi R^2}{4L_m^2 - \pi R^2}.
\label{eq:detailedBal}
\end{equation}

Let $p(x,t)$ and $q(x,t)$ be the probability that a particle is in the passing or blocked state at position $x$ and time $t$, respectively. They obey the coupled set of equations
\begin{equation}
\begin{cases}
    &\partial_t p = - q_{\rm off} p + q_{\rm on} q + D \partial_{xx}p  \\
    &\partial_t q = + q_{\rm off} p - q_{\rm on} q + D \partial_{xx}q  
\label{eq:ratesProblem}
\end{cases}
\end{equation}
with reflection boundary conditions on all walls $\partial_x q|_{x= 0^{\pm}} = 0 \,\, , \,\partial_x p|_{x= \pm L/2} = \partial_x q|_{x= \pm L/2} = 0$. $q$ (but not $p$) is discontinuous in $x = 0$. To infer the probability that, for example, the particle made it to the right starting from the left during time $t$, we can choose initial conditions where the particle is distributed uniformly and in an equilibrium way on the left. This amounts to  $p(x,t=0) =  p_0 \Theta(x < 0)$ and $q(x,t=0) =  q_{\rm off}/q_{\rm on} p(x,t=0)$ where $p_0 = \pi R^2/ 4 L_m^2 L$ and $\Theta$ is the Heaviside function.
The probability that the particle made it to the other side at time $t$ is then  
\begin{equation}
p\left(\int_0^t I(t_1) dt_1 = +1 \right) = \int_0^{L/2} \left[p(x,t)+q(x,t)\right] dx  .
\end{equation}
Finally we obtain the statistics of $\Delta N$ as
\begin{equation}
 \langle    \Delta N^2(t) \rangle =  2 \times 2^2 \times N \times p\left(\int_0^t I(t_1) dt_1 = +1 \right) .
\end{equation}
Framed as such, the \textit{rates model} has similar equilibrium characteristics as the 3D pore. We therefore expect to recover similar noise features and to be able to explain their dependencies. Eventually, we will show that the rates model reproduces exactly the results of 3D Brownian dynamics (BD) simulations. 

\subsection{Solving the rates model}

Analytic solutions to the rates model defined by Eq.~\eqref{eq:ratesProblem} can be found in Laplace space. The method and results are reported in Appendix B. Analytic solutions are in perfect agreement with numerical finite difference solutions of the partial differential equations -- see Fig.~\ref{fig:mapping}-b. We find 
\begin{equation}
    \mathcal{L} \left[ \langle    \Delta N^2 \rangle \right](s) = \frac{4N}{L} \frac{q_{\rm on}}{q_{\rm off}+ q_{\rm on}}  \,\frac{\sqrt{D}}{s^{3/2}} \frac{1 + \frac{q_{\rm on}}{q_{\rm off}}}{\sqrt{\frac{s}{s + q_{\rm on} + q_{\rm off}}} + \frac{q_{\rm on}}{q_{\rm off}}}
    \label{eq:DNLaplace}
\end{equation}
where $\mathcal{L}$ is the Laplace transform, $s$ the Laplace frequency,
$   q = \sqrt{s/D} $ and $\tilde{q} = \sqrt{s/D + (q_{\rm on} + q_{\rm off})/D}$.
Here Eq.~\eqref{eq:DNLaplace} is written at times $t \ll L^2/D$, corresponding to a large box $L$.  Eq.~\eqref{eq:DNLaplace} does not have an analytic form in real time (to the best of our knowledge). However, we may infer limiting behaviors in real time for a few relevant cases. 

\subsection{Narrow pore regime ($R \ll L_m$)}
When $q_{\rm off} \gg q_{\rm on}$ (for the narrow pore $R \ll L_m$) we obtain
an analytic expression in real time as 
\begin{equation}
   \begin{split} \langle    \Delta & N^2(t) \rangle =  \frac{4N}{L} \frac{\pi R^2}{4 L_m^2} \bigg[ \sqrt{\frac{Dt}{\pi}} e^{-(q_{\rm on }+q_{\rm off})t} \,\,  +   \\
       & \frac{\sqrt{D} \left( \frac{1}{2} + (q_{\rm on}+q_{\rm off})t\right) }{\sqrt{q_{\rm on}+q_{\rm off}}}  \mathrm{erf} \left( \sqrt{(q_{\rm on} + q_{\rm off})t} \right) \bigg].
   \end{split} 
\end{equation}

\subsubsection{Early times, $t \ll t_{\rm early} = (q_{\rm off}+ q_{\rm on})^{-1}$.} 
\begin{figure}[h!]
\includegraphics[width = 0.99\columnwidth]{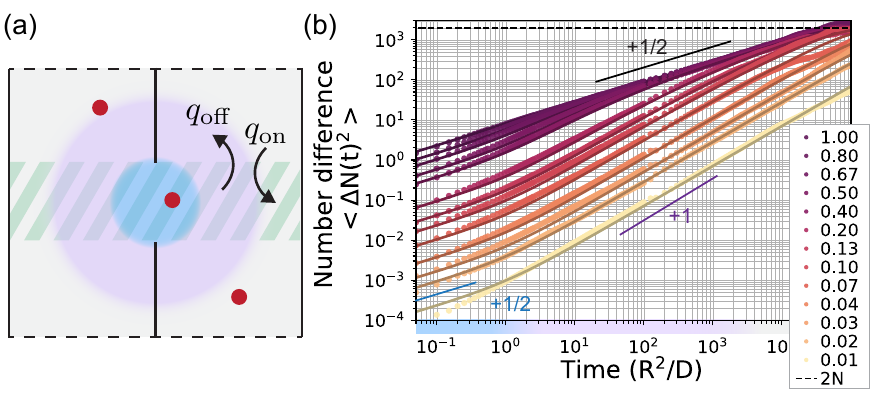}
\caption{\label{fig:correspondance} \textbf{Mechanism for particle fluctuations across a 3D nanopore}. (a) Sketch of the different fluctuation regimes for a small pore: the early regime in blue corresponds to particles located close to the pore mouth; the intermediate regime in purple corresponds to exchanges between the open pore region (dashed green) and the rest of the reservoir; the later regime in gray corresponds to a well mixed situation.  (b) Number difference fluctuations with time for several values of $R/L_m$ as indicated with the various colors. Dots correspond to data from BD and lines to the analytic solution of Eq.~\eqref{eq:ratesProblem}. Matching at very early times and very narrow pores is not perfect due to a lack of significant statistics when only a few particles translocate. The regimes are identified with the same color scheme as in (a). Simulation parameters correspond to Fig.~\ref{fig:fig1}.}
\end{figure}

At very short times $t \ll t_{\rm early} =  (q_{\rm off}+ q_{\rm on})^{-1} \sim R^2/D$ we find (using the circular pore expressions for $q_{\rm on}$ and $q_{\rm off}$):
\begin{equation}
    \langle    \Delta N^2(t) \rangle  \underset{t\ll t_{\rm early}}{=}  \frac{8N}{L} \frac{\pi R^2}{4 L_m^2} \sqrt{\frac{Dt}{\pi}}.
\end{equation}
The fluctuations here are exactly that of a walk on a line corrected by a "geometric" prefactor $\frac{\mathcal{A}_{\rm open}}{\mathcal{A}_{\rm closed}} = \frac{\pi R^2}{4 L_m^2}$ accounting for the open area of the membrane. In fact at early times the particles that participate to the fluctuations are only those that are found very close to the pore -- see blue region in Fig.~\ref{fig:correspondance}-a.  They thus behave exactly as if they were "seeing" no membrane wall -- yet.  This fractional behavior is seen systematically at early times for all pore sizes -- see Fig.~\ref{fig:correspondance}-b.


\subsubsection{Intermediate times, $ t_{\rm early} \ll t \ll t_{\rm int}$.} At intermediate times, namely, when exchanges are now possible between domains in front of the pore and facing the wall $t_{\rm early} \ll t \ll t_{\rm int}$ (the limit $t_{\rm int}$ will be defined in the following paragraph) we find a diffusive regime
\begin{equation}
    \langle    \Delta N^2(t) \rangle  \underset{t_{\rm early} \ll t \ll t_{\rm int}}{=} \frac{4N}{L}  \frac{\pi R^2}{4 L_m^2} \sqrt{D q_{\rm off}} \, t.
    \label{eq:DeltaNnarrowpore}
\end{equation}
Here particle translocation events are dominated by exchanges between the pore region and the rest of the reservoirs -- see purple in Fig.~\ref{fig:correspondance}. Eq.~\eqref{eq:DeltaNnarrowpore} can be further interpreted as an actual random walk. When the opening of the pore is quite small, during the characteristic time  $\tau_{\rm off} = 1/q_{\rm off}$, a small quantity of particles may transition from one side to the other with probability $\delta p = \frac{N}{2} \frac{\delta \mathcal{V}}{\mathcal{V}}$ where $\mathcal{V} = 4 L_m^2 L$ is the total volume and $\delta \mathcal{V}$ is a small volume in front of the pore mouth. This volume writes naturally $\delta \mathcal{V} = \pi R^2 \ell_{\rm off} $ where $\ell_{\rm off} = \sqrt{D/q_{\rm off}}$ is a relevant length scale inside the reservoirs. Particles further than $\ell_{\rm off}$ from the pore mouth will in average not make it through the pore. If that small quantity of particles shifts right, then $\Delta N = +2$ with probability $\delta p$ and -2 with same probability. The steps in $\Delta N$ are uncorrelated; after time $\tau_{\rm off}$, particles in front of the pore have been remixed within the reservoirs and replaced by others. Therefore $\Delta N$ is a Brownian walk with time steps $\tau_{\rm off} $ and step sizes of $\pm$ 2 with probability $\delta p$. We find
\begin{equation}
    \langle  \Delta N^2(t) \rangle =  2^2 \times 2 \delta p \times \frac{t}{\tau_{\rm off} }
\end{equation}
and gathering all quantities we recover Eq.~\eqref{eq:DeltaNnarrowpore}. Such diffusive behavior is also seen systematically at intermediate times for nearly all pore sizes -- see Fig.~\ref{fig:correspondance}-b.

\subsubsection{Later times, $t \gg t_{\rm int}$.} At later times, starting from Eq.~\eqref{eq:DNLaplace} we find in general (regardless of the pore size)  
$ \mathcal{L} \left[  \langle   \Delta N^2 \rangle  \right] (s) \underset{s \rightarrow 0}{=}   \frac{4N}{L}  \frac{\sqrt{D}}{s^{3/2}}$, yielding
\begin{equation}
    \langle    \Delta N^2(t) \rangle \underset{t \gg t_{\rm int}}{=}   \frac{8N}{L} \sqrt{\frac{Dt}{\pi}}.
    \label{eq:fractionalLate}
\end{equation}
At later times, whatever the size of the pore, we recover the bare fractional noise. This later regime corresponds to a phase where the reservoir is now sufficiently well mixed that it doesn't "matter" anymore whether particles are in front of the open pore or not. This later regime may be observed only from times $t \geq t_{\rm int}$, where $t_{\rm int}$ corresponds to the cross-over between the intermediate diffusive Eq.~\eqref{eq:DeltaNnarrowpore} and late fractional noise Eq.~\eqref{eq:fractionalLate}.
This leads to
$
t_{\rm int}   =  \left(\frac{8 L_m^2}{\pi^{3/2} R^2}\right)^2 \frac{1}{q_{\rm off}}.    
$
Note that it is only possible to observe this late time regime if $t_{\rm int} \leq t_{\rm late} \sim \frac{L^2}{D}$, the time when fluctuations reach the $2 N$ limit. 
The later time regime is depicted in Fig.~\ref{fig:correspondance} in gray and is indeed reached for a number of pores.

\subsubsection{Conclusion for narrow pores.}

The number difference $\Delta N$ in narrow pores therefore experiences 4 different phases 
\begin{itemize}
\item fractional noise (as $\sqrt{t}$) for $ t \ll t_{\rm early} \simeq \frac{R^2}{D}$,
\item diffusive noise for  $t_{\rm early} \ll t \ll t_{\rm int} \simeq \frac{L_m^4}{R^4} t_{\rm early}$,
\item fractional noise (as $\sqrt{t}$) for $ t_{\rm int} \ll t \ll t_{\rm late} \simeq \frac{L^2}{D}$,
\item saturation for $t \gg t_{\rm late}$. 
\end{itemize}

Note that here, the time to reach saturation of the fluctuations $t_{\rm late}$ depends in general on the pore size. For very small pore sizes, we can equate Eq.~\eqref{eq:DeltaNnarrowpore} with $2N$ to find that $t_{\rm late} \sim \frac{L_m^2 L}{DR}$. As $t_{\rm late}$ scales inversely with the pore size, this increases the overall time scale over which such regimes may be observed experimentally. 

\subsection{Wide pore regime ($R \sim L_m$)}

When $q_{\rm off} \ll q_{\rm on}$ (for wide pores $R \sim L_m$),
$  \mathcal{L} \left[  \langle   \Delta N^2 \rangle  \right]  (s) =  \frac{4N}{L}  \frac{\sqrt{D}}{s^{3/2}}$
such that we obtain for all times in very broad pores
\begin{equation}
    \langle    \Delta N^2(t) \rangle =  \frac{8N}{L} \sqrt{\frac{Dt}{\pi}} .
\end{equation}
We recover naturally the result for Brownian walk on a line. 


\subsection{Agreement with the 3D pore problem}

To check how the rates model reproduces BD simulations through a 3D pore, we overlap predictions from the analytic solution of the rates model and BD results -- see Fig.~\ref{fig:correspondance}. To complete the mapping we need to specify the value of the rates $q_{\rm on}$ and $q_{\rm off}$. The phenomenological choice 
\begin{equation}
    q_{\rm off}^{-1}= \frac{\pi R^2}{4 D} \,\, \mathrm{and} \,\, q_{\rm on}^{-1} = \frac{4L_m^2 - \pi R^2}{4D}
\end{equation}
obeys detailed balance Eq.~\eqref{eq:detailedBal}. Here we chose the rates as $q = a^2/4D$. $a^2$ corresponds to the characteristic area associated with the rate as expressed in Eq.~\eqref{eq:detailedBal}. The factor $1/4$ corresponds to $1/2d$ where $d = 2$ is the dimension of interest for diffusion parallel to the membrane plane. Such a phenomenological choice accurately reproduces BD simulations -- see Fig.~\ref{fig:correspondance}-a. The agreement is excellent over 6 orders of magnitude in time and for many pore parameters. Fig.~\ref{fig:correspondance} highlights the different regimes (early fractional, intermediate diffusive, later fractional). In particular, at later times for a few intermediate pores ($R \sim 0.2-0.5 L_m$), fractional noise is indeed observed again after a diffusive interval.


Importantly, the rates model is not specific to the circular geometry of the pore and would hold for other geometries such as squares (see Appendix C) or rectangular slits.

\subsection{Noise spectrum in nanopores}

\subsubsection{Noise spectrum}

\begin{figure}[h!]
\includegraphics[width = 0.95\columnwidth]{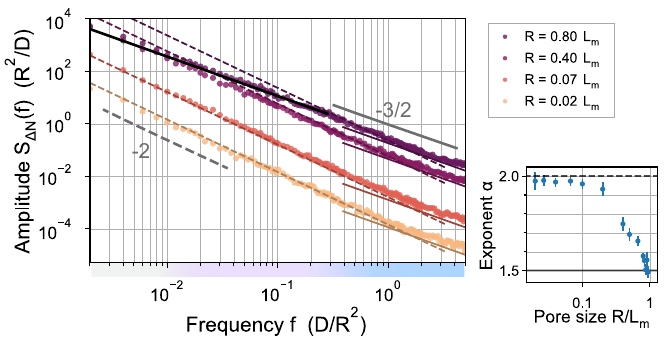}
\caption{\label{fig:spectraPore} \textbf{Noise spectrum in narrow short pores}. Noise spectrum amplitude of the number difference in units of time $R^2/D$ for various pore sizes $R$. Dots correspond to BD data and lines to analytic results for the low frequencies (thick full black line, Eq.~\eqref{eq:SdN1}), intermediate frequencies (dashed, color coded, Eq.~\eqref{eq:SdN2}) and high frequencies (full, color coded, Eq.~\eqref{eq:SdN3}). The different regimes are highlighted with the same color code in the frequency spectrum as in Fig.~\ref{fig:correspondance}. (Inset) Least-squares fit to find power law exponent $\alpha$ in $S_{\Delta N}(f) \sim 1/f^{\alpha}$ over the low frequency range ($f \leq 0.03 D/R^2$). Simulation parameters correspond to that of Fig.~\ref{fig:fig1}.}
\end{figure}

Analysis of the different regimes in time of $\Delta N$ allows us to draw conclusions on the noise spectrum of $\Delta N$ or $I_{\tau}$. Here for simplicity we focus on the noise spectrum properties of $\Delta N$. Similarly as for $S_{\Delta N}$ in the case of the single line we can use the results of Ref.~\onlinecite{burov2011single}
\begin{itemize}
\item at high frequencies (early times) we have 
\begin{equation}
S_{\Delta N}(f) \underset{f \gg t_{\rm early}^{-1}}{=} \frac{\pi R^2}{4 L_m^2}   \frac{2N \sqrt{2D}}{L (2\pi)^{3/2}} \frac{1}{f^{3/2}} = \frac{\pi R^2}{4 L_m^2}    S^{\rm (line)}_{\Delta N}(f) 
\label{eq:SdN1}
\end{equation}
\item at intermediate frequencies
\begin{equation}
S_{\Delta N}(f) \underset{t_{\rm early}^{-1} \gg f \gg t_{\rm int}^{-1}}{=} \frac{\pi R^2}{4 L_m^2} \frac{8N}{L}  \frac{\sqrt{D q_{\rm off}}}{(2\pi)^2}  \frac{1}{f^{2}}\label{eq:SdN2}
\end{equation}
\item at low frequencies (at late times or if the pore size/pore density is large enough)
\begin{equation}
S_{\Delta N}(f) \underset{f\ll  t_{\rm int}^{-1}}{=}   \frac{2N \sqrt{2D}}{L (2\pi)^{3/2}} \frac{1}{f^{3/2}} =  S^{\rm (line)}_{\Delta N}(f). 
\label{eq:SdN3}
\end{equation}
\end{itemize}
The regimes decaying as $1/f^{\alpha}$ with $\alpha = 3/2$ or $2$ are consistently observed in Fig.~\ref{fig:spectraPore} for the variety of pores investigated. The transitions from one regime to another are smooth. As a result, extracting the exponent $\alpha$ over a finite frequency range (see Fig.~\ref{fig:correspondance}, inset) can result in the observation of $\alpha$ values continuously ranging between $1.5$ and $2.0$. This hints that in experimental conditions, where acquisition times are finite, similar real valued exponents could be observed.

\subsubsection{Optimizing signal-to-noise in short pores}

\paragraph{Currents driven by an external field.}

In most nanoporous systems, currents are driven by an external field say $E$. For example, ionic currents are driven by an electric field. Instead of deriving the full consequences of an applied field on the fluctuations, here we make a simple reasoning to infer the expected signal-to-noise in the linear response regime. 

With an applied field we expect to measure an average current of particles scaling as $\langle I_{\rm ext} \rangle = \mathcal{G} E$ where $E$ is the driving force and $\mathcal{G}$ is the pore's conductance, taking into account the geometric parameters of the system. Typically~\cite{secchi2016scaling}, for electric fields we expect $\mathcal{G} \sim R^2$. As a consequence we obtain (within the linear response regime) 
\begin{equation}
\frac{\langle \delta I_{\rm ext}^2 \rangle}{  \langle I_{\rm ext} \rangle^2} = \frac{\langle I_{\rm \tau}^2 \rangle}{\mathcal{G}^2 E^2} \sim \begin{cases}  & \frac{1}{R^2} \,\, \mathrm{for} \,\, f \ll t_{\rm int}^{-1}  \\ 
&\frac{1}{R^4} \,\, \mathrm{for} \,\, f \gg t_{\rm int}^{-1}  \end{cases}
\end{equation}
depending on the range of frequencies under scrutiny. Here $\delta I_{\rm ext} = I_{\rm ext} - \langle I_{\rm ext}\rangle$ corresponds to current fluctuations with respect to the mean and we used the scaling laws in Eqs.~\eqref{eq:SdN1}-\eqref{eq:SdN3}. In this setting we find that wide pores are required to maximize signal-to-noise. 

\paragraph{A note on osmotic currents.}

Though it remains to be assessed in more advanced simulation frameworks (taking explicitly into account the solvent and its differential interaction with the membrane), we expect fluctuations of the number difference, corresponding to the concentration difference, to induce fluctuations in the osmotic pressure drop (at small concentration differences at least), and therefore in osmotic driven currents. We therefore make a short reasoning to infer the signal-to-noise ratio here. When a concentration difference say $\Delta N_0$ is applied between the two pore sides we expect a resulting osmotic current~\cite{marbach2019osmosis} in average as $\langle I_{\rm osm} \rangle \propto \Delta N_0$. Osmotic current fluctuations therefore scale as 
\begin{equation}
\frac{\langle \delta I_{\rm osm}^2 \rangle}{  \langle I_{\rm osm} \rangle^2} = \frac{\langle \Delta N^2 \rangle}{ \Delta N_0^2} \sim \begin{cases}  & \frac{R^2}{L_m^2} \,\, \mathrm{for} \,\, f \ll t_{\rm int}^{-1}  \\ 
&1 \,\, \mathrm{for} \,\, f \gg t_{\rm int}^{-1}  \end{cases} 
\end{equation}
where we made use of the scaling laws in Eqs.~\eqref{eq:SdN1}-\eqref{eq:SdN3}. To maximize signal-to-noise, here, narrow pores, or pores in low density on the membrane, can be used. 

\vspace{2mm}

Overall, in short pores, fractional noise remains predominant at high and low frequencies, while diffusive noise occurs at intermediate frequencies. This results in noise spectral densities scaling as $1/f^{3/2}$ and $1/f^2$. Optimizing signal-to-noise strongly depends on the type of current investigated. Beyond the short pore regime explored here,  long pores (typically at least as long as they are wide) are also common in biological and artificial nanoporous systems.  The purpose of the final section is to identify how fractional noise impacts these long channels. 

\section{Nanochannels, number of "charge carriers", and emergence of $1/f^{1/2}$ noise}
\label{sec:nanochannels}

\begin{figure}[h!]
\includegraphics[width = 0.99\columnwidth]{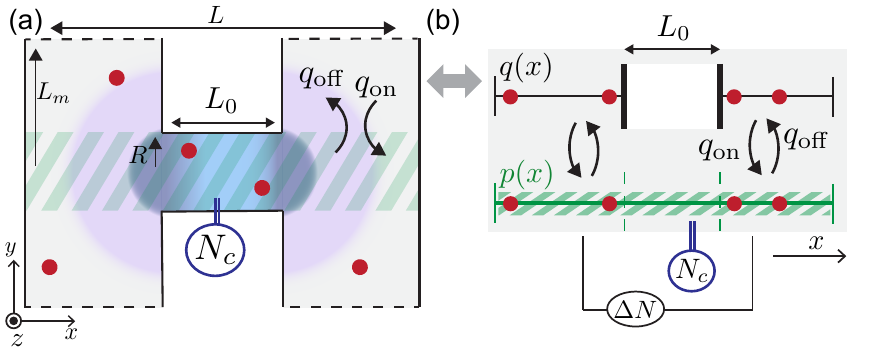}
\caption{\label{fig:ChannelIntro} \textbf{3D nanochannel noise regimes and mapping to a rates model}. (a) 3D nanochannel geometry and different fluctuation regimes for a small pore: the initial regime (dark gray) corresponds to particles located close to the pore mouths;  the early regime (lighter blue) corresponds to particles all along the channel and close to the pore mouths; the intermediate regime (purple) corresponds to exchanges between the open pore region (dashed green) and the rest of the reservoir; the later regime in gray corresponds to a well mixed situation.  (b) Mapping of a 3D nanochannel to a rates model similar to that of Fig.~\ref{fig:mapping}. }
\end{figure}

We now investigate geometries where the pore has a finite length $L_0$ -- see Fig.~\ref{fig:ChannelIntro}. We refer to these systems as \textit{nanochannels} in contrast with \textit{nanopores} which are infinitely short. Importantly, here we may define the number of (uncharged) particles within the pore $N_c$, akin to the "number of charge carriers" -- in analogy with ionic solutions where electric conductance is directly related to the number of charge carriers~\cite{bocquet2010nanofluidics}. We wish to understand what scalings we can expect in the noise, especially for the number of particles within the pore $N_c$. 


\subsection{Very long nanochannels}
\label{sec:veryLongChannels}

As a simplistic yet insightful introduction to nanochannels, we consider "channels" on a  line -- see Fig.~\ref{fig:figChannel1}-a, similar to the treatment of Sec.~\ref{sec:fractionalOrigin}. On this line, the channel is delimited by two imaginary boundaries in $x = \pm L_0/2$ where $L_0$ is the length of the channel. In BD simulations, we still use reflecting boundary conditions in $x = \pm L/2$ to mimic the effect of reservoirs. For simplicity, in analytic derivations we assume $L_0 \ll L$ and we neglect the influence of boundaries.

\subsubsection{Number difference}
\begin{figure}[h!]
\includegraphics[width = 0.99\columnwidth]{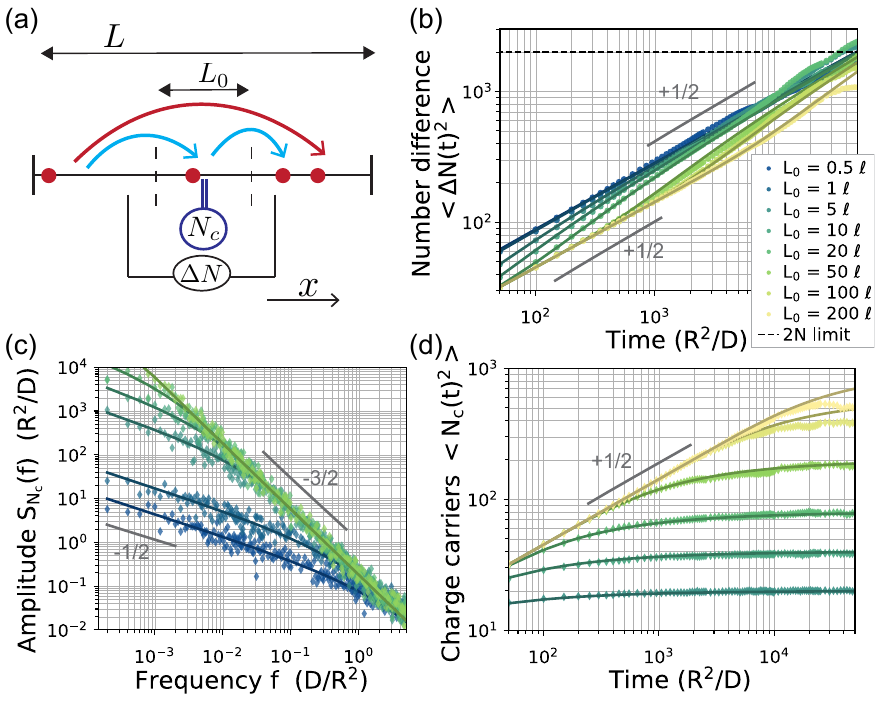}
\caption{\label{fig:figChannel1} \textbf{Fractional noise in long channels}. (a) Illustration of long channels, where we consider particle number difference $\Delta N$ between the left and right sides of imaginary boundaries in $x= \mp L_0/2$ respectively. We also consider the number of particles $N_c$ inside the channel with $ -L_0/2 < x < L_0/2$. Color arrows indicate all jumps to the right contributing to $\Delta N$ or to $N_c$. (b) Mean square number difference $\langle \Delta N^2(t) \rangle$; and (d) number of particles within the pore $\langle N_c^2(t) \rangle$ with time for different values of the channel size $L_0$. Dots are BD data and lines correspond to Eq.~\eqref{eq:DeltaNL0} for (b) and  Eq.~\eqref{eq:NcL0} for (d). Slight mismatch at the largest times can be explained by limited data statistics. (c) Corresponding spectrum $S_{N_c}$ for the number of particles within the channel. Lines correspond to Eq.~\eqref{eq:SNc}. Legend for the size of the channel $L_0$ is shared between panels (b), (c) and (d).   Simulation parameters correspond to that of Fig.~\ref{fig:singleFile}. $N = 1000$ particles were simulated and $\ell$ is an arbitrary length unit. The total simulation time was $1.5 \times 10^{6} \, \ell^2/D$ with a time step $\Delta t = 0.05 \, \ell^2/D$.}
\end{figure}

The relevant number difference corresponds now to the difference in particle number between the right and left sides $\Delta N = N(x > L_0/2) - N(x < -L_0/2)$. Similarly as in Sec.~\ref{sec:fractionalOrigin} this problem is amenable to analytic calculations. The fluctuations of $\Delta N$ are easily expressed in terms of the probability to make jumps from one area to another, with the added complexity that different jumps contribute differently to $\Delta N$. Particles jumping from the left to the right (or inversely) will make a change $+2$ to $\Delta N$ (red arrow) while particles jumping in and out of the channel will only contribute $+1$ to $\Delta N$ (blue arrows) -- see Fig.~\ref{fig:figChannel1}-a. As in Sec.~\ref{sec:fractionalOrigin}, to make analytic derivations simpler, we focus on times $t \ll L^2/D$ such that we can neglect the finite extent of reservoirs. 

We focus on all the jumps towards the right.
The probability distribution of the particle's position, provided it started on the left ($x < -L_0/2$), is 
\begin{equation}
    p _L(x,t) = \frac{\rho_0}{2} \left[ 1 - \erf \left( \frac{x + L_0/2}{\sqrt{4Dt}} \right) \right]
\end{equation}
where $\rho_0 = N/L$ is the concentration of particles. The probabilities to jump from the left (L) to the channel (C), or to the right (R) are
\begin{equation}
    p_{\rm L \rightarrow C}(t) = \int_{-
    \frac{L_0}{2}}^{\frac{L_0}{2}} p_L(x,t) dx\, ; \,\, \,\, p_{\rm L \rightarrow R}(t) = \int_{\frac{L_0}{2}}^{+\infty} p_L(x,t) dx,  
\end{equation}
where here the upper integration bound is $+\infty$ not $L$ as we may neglect the finite extent of reservoirs at short enough times. 

If the particle started in the center, the probability distribution of its position is 
\begin{equation}
    p_C(x,t) = \frac{\rho_0}{2} \left[  \erf \left( \frac{x + L_0/2}{\sqrt{4Dt}} \right) +  \erf \left( \frac{L_0/2-x}{\sqrt{4Dt}} \right) \right]
\end{equation}
and the probability to jump from the center to the right is
\begin{equation}
   p_{\rm C \rightarrow R}(t) = \int_{\frac{L_0}{2}}^{+\infty} p_C(x,t) dx. 
\end{equation}

Finally the fluctuations sum up to
\begin{equation}
\label{eq:setupForN}
    \begin{split}
        \langle \Delta N^2(t) \rangle = & 2 \bigg( 2^2 p_{\rm L \rightarrow R}(t)  + 1^2 p_{\rm L \rightarrow C}(t) + 1^2 p_{\rm C \rightarrow R}(t) \bigg)
    \end{split}
\end{equation}
where the factor 2 in front of the whole expression originates from the fact that particles may jump with equal probability right or left. We stress again that we abbreviate here $ \langle \Delta N^2(t) \rangle = \langle (\Delta N (t) - \Delta N(0))^2 \rangle $ . Standard algebra yields
\begin{equation}
 \label{eq:DeltaNL0}
    \begin{split}
        \langle \Delta N^2(t) \rangle = & 2 N_0 \left[ \left( 1 + e^{-\frac{L_0^2}{4 D t}} \right) \sqrt{\frac{4 D t}{\pi L_0^2}}  -  1 + \erf \left( \frac{L_0}{\sqrt{4 D t}} \right)  \right]
    \end{split}
\end{equation}
where $N_0 = \rho_0 L_0$. Eq.~\eqref{eq:DeltaNL0} corresponds exactly with BD simulations (for $t \ll L^2/D$) -- see Fig.~\ref{fig:figChannel1}-b. This problem may also be solved in Laplace space, with a similar framework as in Appendix B.

Interestingly, the fluctuations of $\Delta N$ feature two relevant limits. At early times we find fractional noise
\begin{equation}
    \langle \Delta N^2(t) \rangle \underset{t \ll L_0^2/D}{=}  4 \frac{N}{L} \sqrt{\frac{D t}{\pi}} 
    \label{eq:DeltaNL0tsmall}
\end{equation}
that is exactly $1/2$ of that observed at longer times
\begin{equation}
    \langle \Delta N^2(t) \rangle \underset{t \gg L_0^2/D}{=}  8 \frac{N}{L} \sqrt{\frac{D t}{\pi}}.
\end{equation}
At long times everything happens as if the channel were infinitely short, as particles have diffused way further than the typical length of the channel. Fluctuations thus are dominated by jumps between the left and right sides. At short times however fluctuations are dominated by particle exchanges from the channel to the reservoirs and \textit{vice versa} -- see dark gray in Fig.~\ref{fig:ChannelIntro}-a. They are similar in nature but contribute twice as less to the fluctuations, therefore explaining the scaling in Eq.~\eqref{eq:DeltaNL0tsmall}. At intermediate times, fluctuations transit from one regime to the other. 

Importantly, for channels fractional noise is preserved. This is clear when one considers again the origin of fractional noise, as explored in Sec.~\ref{sec:fractionalOrigin}. In fact, fractional noise occurs when observing the statistics of random particles transiting from one region to another. For channels, where particles transit from reservoir to pore and pore to reservoir, one thus naturally expects to witness fractional noise. This highlights the universality of fractional noise.

\subsubsection{Number of particles within the pore: "charge carriers".}

In the channel problem, we may also investigate the fluctuations of the number of particles $N_c(t)$ present inside the channel (with $ -L_0/2 < x < L_0/2$). $N_c(t)$ is akin to the number of charge carriers (although here the particles are not charged), that is essential to understand ionic currents in nanoporous systems.~\cite{secchi2016scaling} The average number of particles in the channel is $\langle N_c(t) \rangle = \rho_0 L_0 \equiv N_0$. We can then write the fluctuations as 
\begin{equation}
\langle N_c^2(t) \rangle = 2 \left( 1^2 p_{C\rightarrow R}(t) + 1^2 p_{L\rightarrow C}(t) \right)
\end{equation}
where we abbreviated $\langle (N_c(t) - N_c(0))^2 \rangle = \langle N_c^2(t) \rangle$, and the $1^2$ highlights, as in Eq.~\eqref{eq:setupForN}, the contribution \textit{e.g.} $+1$ to $N_c$ ($1^2$ to $N_c^2$), when a particle goes from left to center. We obtain
\begin{equation}
\begin{split}
\langle N_c^2 (t) \rangle =  & 2 N_0 \left[ \left( 1 - e^{-\frac{L_0^2}{4 D t}} \right) \sqrt{\frac{4 D t}{\pi L_0^2}}  +  1 - \erf \left( \frac{L_0}{\sqrt{4 D t}} \right)  \right].
\end{split}
\label{eq:NcL0}
\end{equation}
This analytic result corresponds exactly with BD -- see Fig.~\ref{fig:figChannel1}-d. The fluctuations of $N_c$ feature two relevant limits. At early times we find fractional noise
\begin{equation}
    \langle N_c^2 (t)   \rangle \underset{t \ll L_0^2/D}{=}  4 \frac{N}{L} \sqrt{\frac{D t}{\pi}} 
    \label{eq:NcL0tsmall}
\end{equation}
corresponding exactly to the early time regime for the number difference, Eq.~\eqref{eq:DeltaNL0tsmall}. In fact, fluctuations of $N_c$ are also dominated by particles in the vicinity of the pore mouth. At longer times the fluctuations plateau
\begin{equation}
    \langle  N_c^2 (t) \rangle \underset{t \gg L_0^2/D}{=}  2 N_0.
\end{equation}
This is naturally expected, for the same reason as a plateau $2N$ is reached at long times for the number difference. 

\subsubsection{Spectrum of the number of particles within the pore.}

As fractional noise is seen also in the number of particles within the pore, we therefore expect its spectrum $S_{N_c}$ to contain a signature of fractional noise as $1/f^{3/2}$. Such behavior is indeed observed for large enough frequencies -- see Fig.~\ref{fig:figChannel1}-c. However, for low frequencies (long times) it is not straightforward to understand the noise spectrum dependence, as the fluctuations $N_c$ saturate. 

Based on the analysis of jumps above and following the method in Ref.~\onlinecite{bezrukov2000particle}, we can completely calculate the frequency spectrum. The correlation function of the number of particles within the pore simply corresponds to the probability that the particle did \textit{not} leave the channel, 
\begin{equation}
\langle N_c(t) N_c(0) \rangle = 1 - 2 p_{C\rightarrow R}(t).
\end{equation}
This can be simply evaluated as
\begin{equation}
\langle N_c(t) N_c(0) \rangle = N_0  \left[  \left( e^{-\frac{L_0^2}{4Dt}} - 1\right) \sqrt{\frac{Dt}{\pi L_0^2}} + \erf \left( \frac{L_0}{\sqrt{4Dt}} \right) \right].
\end{equation}
When the particles have diffused beyond the channel's extent ($t\gg L_0^2/D$), this correlation function decays as $1/\sqrt{t}$ and therefore continues to grow significantly when integrated over time (at least for times $t \ll L^2/D$). As a result we may expect the spectrum at zero frequency to diverge as well -- and not to saturate (as is seen \textit{e.g.} in Ref.~\onlinecite{bezrukov2000particle}). 
To infer the analytic expression for the spectrum, we write the Laplace transform of the correlation
\begin{equation}
\mathcal{L} \left[ \langle N_c(t) N_c(0) \rangle \right](s) =  - N_0\frac{2  \sqrt{D}}{L_0} \frac{1 - e^{-L_0 \sqrt{s/D}}}{s^{3/2}} 
\end{equation} 
and simply
\begin{equation}
S_{N_c}(f) = \mathrm{Real} \left[  \mathcal{L} \left[ \langle N_c(t) N_c(0) \rangle \right](2i \pi f)  \right].
\label{eq:SNc}
\end{equation}
Analytic expansions of Eq.~\eqref{eq:SNc} shows the expected frequency decay at high frequencies 
\begin{equation}
\label{eq:SNcL}
S_{N_c}(f) \underset{f \gg D/L_0^2 }{=} \frac{2 N \sqrt{2 D}}{L} \frac{1}{(2\pi)^{3/2}} \frac{1}{f^{3/2}}.
\end{equation}
At low frequencies, a decay as $1/f^{1/2}$ emerges
\begin{equation}
\label{eq:SNcS}
S_{N_c}(f) \underset{f \ll D/L_0^2 }{=}  N \frac{L_0^2}{\sqrt{2 D} L} \frac{1}{(2\pi)^{1/2}} \frac{1}{f^{1/2}}.
\end{equation}
Both decay laws are consistently obtained in BD simulations, see Fig.~\ref{fig:figChannel1}-c. Eq.~\eqref{eq:SNcS} is quite interesting as it shows the dramatic consequence of the slow decay in the correlation function on the noise spectrum. 
Overall, this demonstrates that low frequency noise is readily observed in our simple system for the number of particles within the pore. Brownian motion is thus sufficient to trigger intriguing noise features, with peculiar $1/f^{\alpha}$ dependence -- without resorting to more complex effects.  

At extremely small frequencies $f \ll D/L^2$, corresponding to very long times $t \gg L^2/D$ the particles feel the finite extent of the reservoirs, and one eventually finds a saturation of the frequency spectrum $S_{N_c}(f) \underset{ }{=}  \frac{N}{8D} \frac{L_0^2}{L^2} \frac{1}{(1-L_0/L)^2}$.
This saturation would be rarely observed in experiments as it would require acquisitions over days (taking \textit{e.g.} $L = 1~\mathrm{cm}$ and $D = 2 \times 10^{-9}~ \mathrm{m^2/s}$) and therefore is not relevant in general. Note that this saturation is not comparable to the saturation observed in Ref.~\onlinecite{bezrukov2000particle}, that examines infinite reservoirs. 

\vspace{2mm}

When considering realistic geometries, with a thick membrane, we may therefore expect different noise regimes according to the relative values of the pore width (radius, or typical cross section size) and the pore length. The following section is dedicated to summarizing the scalings and transitions between behaviors in the general nanochannel geometry, with a focus on the number of particles within the pore. 

\subsection{General geometry and consequences for charge carrier fluctuations}

\subsubsection{Mapping to a rates problem and limit regimes}

Similarly as for the narrow problem, we can map the nanochannel problem to a rates problem -- see Fig.~\ref{fig:ChannelIntro}-b. We use similar notations and take $p(x,t)$ and $q(x,t)$ the probability that a particle is in the passing or blocked state at position $x$ and time $t$. Compared to the short pores in Sec.~\ref{sec:narrowPores}, here, over the channel length $L_0$, the blocked state does not exist; the passing state does not exchange with the blocked state. As in Sec.~\ref{sec:narrowPores}, we can solve for the probability distribution functions $p$ and $q$ in Laplace space and then calculate the fluctuations of $\Delta N$ and $N_c$ using a similar formalism as for Eq.~\eqref{eq:setupForN}. Full solutions are detailed in Appendix D.

The results (both analytic and of BD simulations) point, as expected, to an interplay of channel-like behavior at short times as seen in Sec.~\ref{sec:veryLongChannels} and pore-like behavior at longer times as in Sec.~\ref{sec:narrowPores}. We can distinguish 5 phases (considering pores that are at least as long as they are wide $L_0 \gtrsim R$):
\begin{itemize}
\item (initial, dark gray in Fig.~\ref{fig:ChannelIntro}-a) For $ t \ll t_{\rm early} = \frac{R^2}{D}$: fractional noise ($ \langle \Delta N(t)^2 \rangle = \langle N_c(t)^2 \rangle = 4 c_0 \mathcal{G_R} \sqrt{D t}$) where $\mathcal{G_R} $ is a geometric prefactor taking into account the details of the channel geometry and $c_0$ is the average particle concentration. 
\item (early, light blue in Fig.~\ref{fig:ChannelIntro}-a) For $ t_{\rm early} \ll t \ll t_{\rm channel} = L_0^2/D$, fractional noise with twice as large amplitude ($ \langle \Delta N(t)^2 \rangle = \langle N_c(t)^2 \rangle = 8 c_0 \mathcal{G_R} \sqrt{D t}$) 
\item  (intermediate, purple in Fig.~\ref{fig:ChannelIntro}-a) For $t_{\rm channel} \ll t \ll t_{\rm int} \simeq \frac{L_m^4}{R^4} R^2/D$ diffusive noise in the number difference; saturation for $N_c$.
\item (later, gray in Fig.~\ref{fig:ChannelIntro}-a) For $ t_{\rm int} \ll t \ll t_{\rm late} \simeq \frac{L^2}{D}$ fractional noise in the number difference, ($ \langle \Delta N(t)^2 \rangle = 8 \rho_0 \sqrt{D t}$); saturation for $N_c$.
\item (final) For $t \gg t_{\rm late}$, saturation for all variables. 
\end{itemize}
Full derivations and agreement of BD simulations with analytic results showing the interplay of these 5 regimes are reported in Appendix D in Figs.~\ref{fig:appendix2} and~ \ref{fig:appendix3}. We now turn to the investigation of the noise spectrum. 

\subsubsection{Spectrum of the number of particles within the pore.}

\begin{figure}[h!]
\includegraphics[width = 0.99\columnwidth]{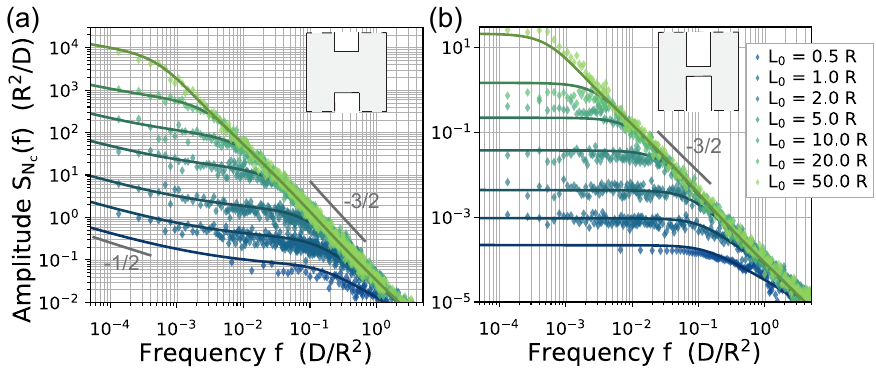}
\caption{\label{fig:MappingChannel} \textbf{Noise on the number of particles within the pore in various pores}. Noise spectrum of $N_c$ for (a) broad pores with $R = 0.4 L_m$ and (b) narrow pores with $R = 0.02 L_m$ for different pore lengths as indicated in the shared legend. Dots are results from BD simulations. Here the pore is square with side $2R$. Full lines correspond to the analytic result of Eq.~\eqref{eq:CorrNc3D}. Other simulation parameters correspond to that of Fig.~\ref{fig:fig1}.}
\end{figure}

\paragraph{Noise spectrum.}
Following a similar approach as in Sec.~\ref{sec:veryLongChannels}, we find the Laplace transform of the correlation function
\begin{equation}
\begin{split}
\mathcal{L} & \left[ \langle N_c(t) N_c(0) \rangle \right](s) =  - N_0\frac{4  \sqrt{D}}{L_0} \frac{1}{s^{3/2}} \left( 1 - e^{-q L_0 } \right) \times \\
& \,\,\,\,\, \,\,\, \frac{\tilde{q} (q_{\rm on} + q_{\rm off})}{e^{-q L_0 } q_{\rm off}(\tilde{q} - q) + q q_{\rm off}+(q_{\rm off}+2 q_{\rm on}) \tilde{q}}
\end{split}
\label{eq:CorrNc3D}
\end{equation} 
where we recall that $q = \sqrt{s/D}$ and $\tilde{q} = \sqrt{q^2 + (q_{\rm on} + q_{\rm off})/D}$. Here $N_0 = \langle N_c(t) \rangle = c_0 \pi R^2 L_0$ is the average number of particles within the pore and $c_0$ the particle concentration.  Using Eq.~\eqref{eq:SNc} we can fully obtain the spectrum of fluctuations for $N_c$ that agrees remarkably with BD simulations -- see Fig.~\ref{fig:MappingChannel}. 

Expanding Eq.~\eqref{eq:CorrNc3D} it is possible to obtain limiting relevant regimes for the fluctuation spectrum. Here we assume that the channel is rather isolated on the membrane $L_m \gtrsim L_0$ (corresponding to $q_{\rm on} \ll D/L_0^2$), such that 3 characteristic behaviors emerge in the fluctuation spectrum:  

\vspace{2mm}

$\bullet$ at very low frequencies, the $1/f^{1/2}$ decay is observed
\begin{equation}
\label{eq:SNcS3D}
S_{N_c}(f) \underset{f \ll q_{\rm on} }{=} \frac{q_{\rm on}}{q_{\rm on}+q_{\rm off}} N_0\frac{L_0}{\sqrt{2 D}} \frac{1}{(2\pi)^{1/2}} \frac{1}{f^{1/2}}.
\end{equation}
which is exactly Eq.~\eqref{eq:SNcS} multiplied by a geometric prefactor $\frac{q_{\rm on}}{q_{\rm on}+q_{\rm off}} = \frac{\pi R^2}{4L_m^2}$ corresponding to the open pore area (for a circular pore of radius $R$). As mentioned earlier, this regime is not seen in theoretical derivations of Ref.~\onlinecite{bezrukov2000particle}. In fact Ref.~\onlinecite{bezrukov2000particle} assumes an infinitely small pore on a membrane, giving $q_{\rm on} \rightarrow 0$, and hence the $1/f^{1/2}$ regime does not appear.  

\vspace{2mm}

$\bullet$ at intermediate frequencies we find a plateau 
\begin{equation}
S_{N_c}(f) \underset{q_{\rm on} \ll f \ll L_0^2/D }{=} \frac{2}{3} N_0 \frac{L_0^2}{D} \left( 1 + \frac{6}{L_0} \sqrt{\frac{D}{q_{\rm off}}}\right).
\label{eq:SdNc3Dplateau}
\end{equation}
This (temporary) saturation corresponds to intermediate times where no significant exchange with the reservoirs is possible yet. Eq.~(15) of Ref.~\onlinecite{bezrukov2000particle} also predicts a saturation for the circular pore $S_{N_c}(f) = \frac{1}{3} N_0 \frac{L_0^2}{D} \left( 1 + \frac{3 \pi}{2} \frac{R}{L_0}\right)$ while we find $S_{N_c}(f) = \frac{2}{3} N_0 \frac{L_0^2}{D} \left( 1 + \frac{3 }{\sqrt{\pi}} \frac{R}{L_0}\right)$. These slight differences originate from the approximate model for reservoirs in Ref.~\onlinecite{bezrukov2000particle}.

\vspace{2mm}

$\bullet$ at high frequencies we recover the $1/f^{3/2}$ decay
\begin{equation}
S_{\Delta N}(f) \underset{f \gg D/L_0^2 }{=} \frac{2 N_0}{L_0} \sqrt{2 D}  \frac{1}{(2\pi )^{3/2}} \frac{1}{f^{3/2}}\label{eq:SdNc3Dlong}.
\end{equation}
This decay corresponds exactly to that observed in infinitely long channels Eq.~\eqref{eq:SNcL}.
Ref.~\onlinecite{bezrukov2000particle} predicts a similar regime for small pores (with an amplitude twice as large) yet only at intermediate frequencies ($f \ll D/R^2$). At large frequencies Ref.~\onlinecite{bezrukov2000particle} finds a decay as $1/f^2$. These differences originate from the approximate model for reservoirs in Ref.~\onlinecite{bezrukov2000particle}. A graphical comparison with Ref.~\onlinecite{bezrukov2000particle} is reported in Appendix D in Fig.~\ref{fig:figBeru}.

In contrast with very long channels, \textit{actual} channels feature a plateau in the frequency spectrum. The extent of this plateau in the frequency spectrum is longer if the channel is more narrow. For example, the onset of the $1/f^{1/2}$ regime occurs at much smaller frequencies for the long channel of Fig.~\ref{fig:MappingChannel}-b than for the short channel of  Fig.~\ref{fig:MappingChannel}-a (hence it is not seen in Fig.~\ref{fig:MappingChannel}-b). Other numerical setups also observe the emergence of such a plateau.~\cite{gravelle2019adsorption}

\vspace{2mm}

\paragraph{Optimizing signal-to-noise}
We now discuss how our results can be harnessed to optimize signal-to-noise for currents related to the number of the number of particles within the pore. Similarly as in Sec.~\ref{sec:narrowPores}, if an external field $E$ is applied we expect to measure an average particle current scaling as $\langle I_{\rm ext} \rangle = \langle \mathcal{G} \rangle E$ where $\mathcal{G}$ is the pore's conductance. Yet conductance (at large enough concentrations, for electric fields) is proportional to the number of charge carriers~\cite{secchi2016scaling}. Although our particles are uncharged, we expect that \textit{e.g.} for strong electrolytes and systems where charge effects are not predominant, the results uncovered for uncharged particles would translate for charged species. We therefore use $N_c$ as a representative for the number of charge carriers,
\begin{equation}
\frac{\langle \delta I_{\rm ext}^2\rangle}{\langle I_{\rm ext}\rangle^2} =  \frac{\langle \delta G^2\rangle}{\langle G\rangle^2}  = \frac{\langle N_c^2 \rangle}{N_0^2} \sim \begin{cases} &  1/c_0 L_m^2 \,\, , \,\,  \mathrm{for} \,\, f \ll q_{\rm on} \\ & L_0^2/N_0   \,\, , \,\,  \mathrm{for} \,\,q_{\rm on} \ll f \ll  D/L_0^2 \\ &  1/N_0  \,\, , \,\,   \mathrm{for} \,\,f \gg  D/L_0^2\end{cases}
\end{equation}
The signal-to-noise ratio can thus be maximized working at high concentrations (large $c_0$ and $N_0$) in quite short pores ($L_0$). Narrow pores (small $R$) decrease the signal to noise ratio (by decreasing $N_0$) and therefore short pores are preferred in this setting where noise originates from purely diffusive mechanisms.  

\section{Conclusions and discussion}

\paragraph*{Emergence of fractional noise in nanoporous systems.}
Fractional noise originates from particle exchanges between one region to another. In nanoporous systems, such exchanges are ubiquitous as solute particles go from reservoir to pore and pore to reservoir. As a consequence, fractional noise emerges naturally in nanoporous systems.  It yields fluctuations in time scaling as $\sqrt{t}$ and traces in the low-frequency noise spectrum decaying as $1/f^{3/2}$. We have demonstrated the presence of such low-frequency traces in various pore geometries. Brownian motion is thus already a key ingredient to trigger such low frequency noise -- without resorting to more complex effects.  

Such $1/f^{3/2}$ dependence in the noise spectrum has been consistently seen in many different settings -- though never rationalized as a generic feature inheriting from fractional noise. In artificial systems, $1/f^{3/2}$ has been measured~\cite{wen2017generalized} and also $1/f^2$ in narrow pores.~\cite{bezrukov2000examining,siwy2002origin} In more advanced numerical systems (including adsorption in the inner pore as compared to our simulations), $1/f^{3/2}$ dependencies have been seen in the number of particles present within the pore.~\cite{gravelle2019adsorption} In approximate theoretical models, $1/f^{3/2}$ and $1/f^{2}$ have been consistently seen as well~.\cite{bezrukov2000particle,zevenbergen2009electrochemical,krause2014brownian}
These results point to the fact that fractional noise decaying as $1/f^{\alpha}$ with $\alpha = 1.5 - 2$ prevails in real systems and indeed has consequences even when more complex effects are at play. 

We also discussed a $1/f^{1/2}$ dependence in the noise spectrum for the number of particles within the pore (akin to the number of charge carriers for charged particles), appearing for frequencies $f \ll L_m^2/D$.  Similarly, low frequency noise decaying as $1/f^{\alpha}$ with $\alpha \simeq 0.5$ has also been observed in a few experimental or numerical systems~.\cite{powell2009nonequilibrium,gravelle2019adsorption} However, it is harder to speculate that fractional noise is the origin of such  measurements. In fact, it involves very low frequencies, either not attainable experimentally or were a number of other processes may very well be at play (such as adsorption/desorption.~\cite{gravelle2019adsorption})


%
%
%


\paragraph*{The effect of reservoirs can be reduced to a 1 dimensional rates problem.}

In this work we have introduced a method to map a complex 3D geometry with reservoirs and pores, to a simple 1D geometry (or more precisely $2 \times 1$D). This mapping relies on transition rates from the passing to the blocked lines and \textit{vice versa}. These rates are established from detailed balance equilibrium and do not rely on any additional assumption. Remarkably, such a mapping allows to reproduce with perfect accuracy the results of 3D simulations. It also opens perspectives to drastically simplify numerical simulations (by simulating particles on 1D lines instead of 3D reservoirs) and analytic calculations. 

Note, that this is in sharp contrast with other theoretical investigations, that rely either on approximate rates of entrance/exit in the pore~\cite{bezrukov2000particle,zevenbergen2009electrochemical}, or on approached, simplified geometries.~\cite{gravelle2019adsorption} It also allows to probe efficiently the effect of different pore geometries.~\cite{yilmaz2021role}

The mapping has great potential to reduce the cost of simulating large reservoirs and probe further effects on nanoporous transport. For example, the mapping could easily be extended to the investigation of more varied geometries, adsorption within the pore~\cite{gravelle2019adsorption} or equilibrium reactions at boundaries mimicking electrodes. Its applicability to other systems remains to be assessed. For example, when charges are added, and electric fields may affect the motion of ions significantly between regions, such a mapping may have to be adapted.

%
%
%



\paragraph*{Further discussion.}

In essence, fractional noise is expected to occur in many systems beyond nanoporous transport. For example, such fractional noise or $1/f^{3/2}$ has been observed in the context of electrochemistry at surfaces.~\cite{macfarlane1950theory,van1965fluctuation} Furthermore, fractional Brownian walks have been used to model or explain subdiffusion patterns for molecules such as mRNA or other large molecules evolving in crowded environments such as cells~\cite{deng2009ergodic, tabei2013intracellular} or with adsorption to surfaces.~\cite{fernandez2020diffusion} It remains to be assessed whether such fractional behavior originates from the same physical principles (namely particles transitioning between one region and another) or from other mechanisms. 

Interestingly, our study shows also how crucial the parameters of the experimental measurement may affect observation. For example, fitting of the low noise frequency over only a few decades may lead to a variety of decay exponents (as was observed in Fig.~\ref{fig:correspondance}-b). Furthermore, currents depending on the acquisition frequency may experience more or less noise. Acquisition frequency dependency of nanopore conductance has been measured in specific cases.~\cite{rauh2017extended}
%

Beyond this equilibrium context, it remains to be assessed how fractional noise survives out-of-equilibrium. For example, we can expect the probability of events where particles exchange back and forth from the pore to the reservoir to decay with an applied external field. This may result in a non-linear dependence of the noise with applied external field. To some extent this is reminiscent -- although implying a different mechanism -- of other non-linearities depending on applied field, for example in conductivity measurements of charged solutions.~\cite{lesnicki2020field} 

%
%

%

%

%

\section*{Acknowledgments}

S.M. is indebted to Aleksandar Donev for acute scientific advice and numerous discussions. S.M. recognizes the help of Michel Pain who pointed to relevant references in the pure math literature associated with this problem. S.M. is further thankful to Alejandro L. Garcia, Benjamin Rotenberg and Miranda Holmes-Cerfon for fruitful discussions. S.M. was supported in part by the MRSEC Program of the National Science Foundation under Award Number DMR-1420073. S.M. acknowledges funding from a Marie-Sklodowska Curie fellowship award number 839225 -- MolecularControl project.   

The present article has been submitted to Journal of Chem- ical Physics. After it is published, it will be found at https://publishing.aip.org/

\section*{Appendix}

\subsection*{Appendix A}

\subsubsection*{A.1. Brownian dynamics simulations}

Brownian dynamics of particles translocating through pores are implemented using a custom made Python routine.

\paragraph{System parameters.} The simulation is performed in non-dimensional time and length scales. The length scale of reference is set to be the pore size $R$ and the time scale of reference $\frac{R^2}{D}$. In general the time step was taken to be $\Delta t = 0.05 \frac{R^2}{D}$ and is much smaller compared to the smallest time scale of the system $\frac{R^2}{D}$. In general $N = 1000$ particles were simulated over $3\times 10^7$ time steps. Other system parameters (such as box size $L$ and membrane size $L_m$) are always specified in figure captions where the relevant data is shown.

In systems made of particles on a line, where no pore size is defined, we use $\ell$ as the unit length and $\ell^2/D$ as the unit time.  

\paragraph{Dynamics.} The noise generation is done through Python's numpy random number generator. At each time step, for each particle, a random number with standard normal distribution (numpy.random.randn) is generated and the particle's position is updated via Eq.~\eqref{eq:starting}. For each independent simulation, the seed is set to a different value. Reflections on the reservoir walls and on the membrane are implemented using De Michele's algorithm~\cite{scala2007event}. All measured quantities were thoroughly checked to be independent of simulation parameters. In particular, they were checked to be independent of the time step $\Delta t$. 

We stress that the particles are intended to be the most simple brownian walkers. They are non-interacting and point-like particles. 

\subsubsection*{A.2. Partial differential equation solvers}

To solve numerically the rates problem defined by Eq.~\eqref{eq:ratesProblem}, a  standard forward Euler scheme was implemented in a custom made Python routine. The numerical solutions were independent of the chosen time and space step. In general the time step used was $\Delta t = 0.05 - 0.005 \, \frac{R^2}{D}$ (increasing to larger values after smoothing out of the initial step functions) and space step $\Delta x = L/(N_x-1)$ with $N_x = 2000$. With $L = 500 R$ in general, this gives $\Delta x = 0.25 R$ and therefore $\Delta t \ll \Delta x^2/D$ is always verified, ensuring the stability of the forward Euler scheme.

\subsection*{Appendix B: Solving the rates problem}

To solve the rates problem defined by Eq.~\eqref{eq:ratesProblem} we focus on the eigenvectors of the partial differential equation system
\begin{equation}
    \begin{cases}
    & z_1 = p + q \\
    & z_2 = q_{\rm off} p - q_{\rm on} q
    \end{cases}
\end{equation}
that obey an uncoupled system of equations
\begin{equation}
\begin{cases}
    & \partial_t z_1 = D \partial_{xx} z_1 \\
    & \partial_t z_2 = - (q_{\rm off} + q_{\rm on}) z_2 + D \partial_{xx} z_2 \\
    & \partial_x z_1|_{x= \pm L/2} = \partial_x z_2|_{x= \pm L/2} = 0
\end{cases}
\end{equation}
To specify the boundary conditions we write $z_i^L = z_i(x\leq 0)$ and $z_i^R = z_i(x\geq 0)$ the left of the wall and right of the wall components of the eigenvectors. They verify
\begin{equation}
\begin{cases}
    & \partial_x z_1^L(0,t) = \partial_x z_1^R(0,t) \\
    & \partial_x z_2^L(0,t) = \partial_x z_2^R(0,t) \\
    & z_1^L(0,t) q_{\rm on} + z_2^L(0,t) = z_1^R(0,t) q_{\rm on} + z_2^R(0,t) \\
    & \partial_x z_1^L(0,t) q_{\rm off} =  \partial_x z_2^L(0,t)
\end{cases}
\end{equation}
and $z_1$ and $z_2$ are both discontinuous in $0$. The initial conditions verify
\begin{equation}
    \begin{cases}
        & z_2^L(x< 0,0) = z_2(x>0,0) = 0 \\
        & z_2^L(0,0) = q_{\rm off} p_0/2 - q_{\rm on} p_0 \frac{q_{\rm off}}{q_{\rm on}} = - q_{\rm off} p_0/2 \\
        & z_2^R(x=0,0) = -z_2^L(0,0) = q_{\rm off} p_0/2 \\
        & z_1^L(x<0, 0) = p_0 \frac{q_{\rm on} + q_{\rm off}}{q_{\rm on}}   \\
        & z_1^L(x>0,0) = 0 \\
        & z_1^L(x=0,0) = p_0 /2 + p_0 \frac{q_{\rm off}}{q_{\rm on}} = p_0 \frac{q_{\rm on} + 2q_{\rm off}}{2q_{\rm on}} \\
        & z_1^R(x=0,0) = p_0 / 2 + 0 =  z_1^L(x=0,0) - 2 p_0 \frac{q_{\rm off}}{2q_{\rm on}}.
    \end{cases}
\end{equation}
Importantly, this set of equations and boundary conditions is compatible with a wrapping
\begin{equation}
\begin{cases}
    & z_2^R(x) = \tilde{z_2}(x) \\
    & z_2^L(-x) = - \tilde{z_2}(x) \\
    & z_1^R(x) =  \tilde{z_1}(x) \\
    & z_1^L(-x) = 2z_0 - \tilde{z_1}(-x) 
\end{cases}
\end{equation}
where $z_0 = p_0 \frac{q_{\rm on} + q_{\rm off}}{2 q_{\rm on}}$, $\tilde{z_i}$ need only be defined on a half space and the boundary conditions are
\begin{equation}
    \begin{cases}
     & \tilde{z_1}(x>0,0) = 0, \tilde{z_1}(x=0,0) = p_0/2 \\
     & \tilde{z_2}(x>0,0) = 0, \tilde{z_2}(x=0,0) = q_{\rm off} p_0/2 \\
     & (z_0 - \tilde{z_1}(0,t)) q_{\rm on} = \tilde{z_2}(0,t)  \\
     & \partial_x \tilde{z_1}(0,t) q_{\rm off} =  \partial_x \tilde{z_2}(0,t).
    \end{cases}
\end{equation}
These boundary conditions are well suited for solving the problem in Laplace space. We define the Laplace transform
\begin{equation}
   \hat{z}_i(x,s) = \int_0^{\infty} e^{-st} \tilde{z}_i(x,t) dt
\end{equation}
where we use here for compactness the notation $\hat{z}$ for the Laplace transform instead of $\mathcal{L}[z]$. The general solution to the system of equations is 
\begin{equation}
    \hat{z}_i = A_i e^{q_i x} + B_i e^{-q_i x}
\end{equation}
where $q_1 = q = \sqrt{\frac{s}{D}}$ and $q_2 = \tilde{q} = \sqrt{q^2 + \frac{q_{\rm off}+ q_{\rm on}}{D}}$. And the integration constants $A_i, B_i$ are determined by the boundary conditions and obey the set of equations
\begin{equation}
    \begin{cases}
     & A_1 e^{qL/2} - B_1 e^{-qL/2} = 0 \\ 
     & A_2 e^{\tilde{q}L/2} - B_2 e^{-\tilde{q}L/2} = 0 \\ 
     & (z_0 \frac{1}{s} - A_1 - B_1) q_{\rm on} = A_2 + B_2 \\
     & (A_1 - B_1) q q_{\rm off} = (A_2 - B_2)\tilde{q}
    \end{cases}
\end{equation}
We recall that 
\begin{equation}
 \langle    \Delta N^2(t) \rangle =  4N \int_0^{L/2} \left[p(x,t)+q(x,t)\right] dx     
\end{equation}
such that
\begin{equation}
    \begin{split}
         \partial_t \langle    \Delta N^2(t) \rangle &=  4N\int_0^{L/2} \partial_t\left[p(x,t)+q(x,t)\right] dx   \\
         &=  4N\int_0^{L/2} D\partial_{xx}\left[p(x,t)+q(x,t)\right] dx    \\
         &= - 4N\, D\partial_x\left[p(x,t)+q(x,t)\right]_{x=0} \\
         &= - 4N \,D\partial_x\left[z^R_1\right]_{x=0} 
    \end{split}
\end{equation}
We can thus obtain in Laplace space
\begin{equation}
    \langle    \hat{\Delta N^2}(t) \rangle = - \frac{4N}{q} (A_1 - B_1) 
\end{equation}
The resulting $\langle    \hat{\Delta N^2}(t) \rangle$ is 
 \begin{equation}
    \langle    \hat{\Delta N^2}(t) \rangle = 8N\, z_0 \frac{\sqrt{D}}{s^{3/2}} \frac{q_{\rm on}}{q_{\rm off}\frac{q}{\tilde{q}} \mathrm{coth} \left( \tilde{q}L/2\right) + q_{\rm on} \mathrm{coth} \left( qL/2\right)} 
\end{equation}
for which no analytic real time expression exists. 

\paragraph*{Analytic result at long times}

Fluctuations at long times correspond in Laplace space to small values of $s$. We find for those the asymptotic behavior
\begin{equation}
    \langle    \hat{\Delta N^2}(t) \rangle \underset{s\rightarrow 0}{=} 4 N\, z_0 \frac{\sqrt{D}}{s^{3/2}} q L = 4 N\, z_0 L \frac{1}{s} 
\end{equation}
translating to real time and making use of the expression of $z_0$ we obtain
\begin{equation}
    \langle    \Delta N^2(t) \rangle \underset{t\rightarrow \infty}{=} = 2N
\end{equation}
which is exactly what is expected. 

\subsection*{Appendix C: Data for the square pore}

\begin{figure}[h!]
\includegraphics[width = 0.85\columnwidth]{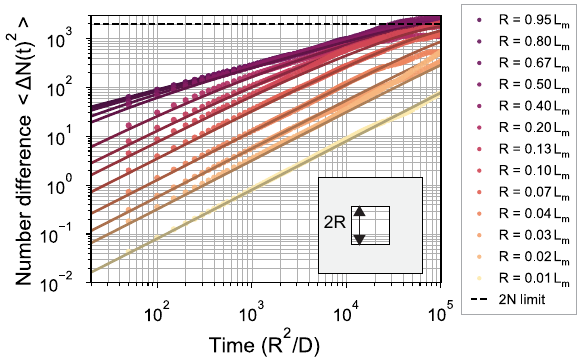}
\caption{\label{fig:mappingSquare} \textbf{The rates model reproduces the regimes observed in a 3D square pore}. Number difference fluctuations with time for several values of $R/L_m$ as indicated with the various colors. Dots correspond to data from BD simulations and lines to the analytic solutions of Eq.~\eqref{eq:ratesProblem}. Other numerical parameters correspond to that of Fig.~\ref{fig:correspondance}-b.}
\end{figure}

For a square pore of side $2R$ we use the phenomenological rates
\begin{equation}
    q_{\rm off}^{-1} = \frac{(2R)^2}{4 D} \,\, \mathrm{and} \,\, q_{\rm on}^{-1} = \frac{4L_m^2 - (2R)^2}{4D}.
\end{equation}
In Fig.~\ref{fig:mappingSquare}, we present the fluctuations $\langle \Delta N(t)^2 \rangle$ with time using the rates model with this phenomenological choice and performing BD simulations through a square pore. The model and numerical data are in perfect agreement. 

\subsection*{Appendix D: Solving the nanochannel problem}

\subsubsection*{D. 1. Rates problem in the nanochannel geometry}
We use similar notations and take $p(x,t)$ and $q(x,t)$ as the probability that a particle is in the passing or blocked state respectively at position $x$ and time $t$. Those quantities obey the coupled set of equations
\begin{equation}
\begin{cases}
    &\mathrm{for} \,\, |x| > L_0/2 , \, \displaystyle\begin{cases} & \partial_t p = - q_{\rm off} p + q_{\rm on} q + D \partial_{xx}p  \\
    &\partial_t q = + q_{\rm off} p - q_{\rm on} q + D \partial_{xx}q  
    \end{cases} \\
    &\mathrm{for} \,\, |x| \leq L_0/2 \,\, , \, \partial_t p =  D \partial_{xx}p  \\
    &\partial_x q|_{x= \pm L_0/2} = 0 \, \partial_x p|_{x= \pm L/2} = \partial_x q|_{x= \pm L/2} = 0
\label{eq:ratesProblemChannel}
\end{cases}
\end{equation}
and $q$ is not defined for $-L_0 < 2 x < L_0$. $p$ and $\partial_x p$ however are continuous over the whole domain, especially in $x = \pm L_0/2$. 
To calculate the fluctuations of $\Delta N$ we must proceed as above and consider several jumps, starting from either the left hand side or the channel itself. Here we report as an example the initial conditions for the particle starting on the left (taking the initial values corresponding to the probabilities to find the particle on the left at equilibrium)
\begin{equation}
    \begin{cases}
    & p(x< -L_0/2,t=0) =  \frac{\pi R^2}{\mathcal{V}} = p_0\\
    & q(x<-L_0/2,t=0) =  \frac{(4L_m^2 - \pi R^2) }{\mathcal{V}} = p_0 \frac{q_{\rm off}}{q_{\rm on}}\\
    & p(x > -L_0/2,t=0) = q(x >-L_0/2,t=0) = 0. 
\end{cases}
\end{equation}
Here $\mathcal{V} =4L_m^2  (L-L_0) + L_0 \pi R^2$ is the total accessible volume. Similarly as in the very long nanochannel problem, we now have to distinguish between particles that actually went to the other side, and particles that only made it inside the channel. 
The probability that the particle made it to the other side at time $t$ is  
\begin{equation}
p_{L\rightarrow R}(t) =  \int_{L_0/2}^{L/2} \left[p(x,t)+q(x,t)\right] dx     
\end{equation}
and the probability that it went inside the channel is
\begin{equation}
p_{L \rightarrow C}(t) =  \int_{0}^{L_0/2} p(x,t) dx .        
\end{equation}
A similar formalism applies for particles starting inside the channel. 

\subsubsection*{D. 2. Solving the rates problem}

To solve the rates problem defined by Eq.~\eqref{eq:ratesProblemChannel}, we adopt a similar method as in appendix B and focus on the eigenvectors of the partial differential equation system. Beyond the channel, for $|x| > L_0/2$ we take
\begin{equation}
    \begin{cases}
    & z_1 = p + q \\
    & z_2 = q_{\rm off} p - q_{\rm on} q
    \end{cases}
\end{equation}
and inside the channel we simply take
\begin{equation}
    z_m = p
\end{equation}
that obey an uncoupled system of equations
\begin{equation}
\begin{cases}
    & \partial_t z_1 = D \partial_{xx} z_1 \,\, \mathrm{for } |x| > L_0/2 \\
    & \partial_t z_m = D \partial_{xx} z_m \,\, \mathrm{for } |x| \leq L_0/2 \\
    & \partial_t z_2 = - (q_{\rm off} + q_{\rm on}) z_2 + D \partial_{xx} z_2 \,\, \mathrm{for } |x| > L_0/2 \\
    & \partial_x z_1|_{x= \pm L/2} = \partial_x z_2|_{x= \pm L/2} = 0.
\end{cases}
\end{equation}
To specify the remaining boundary conditions, we split the domain and write $z_i^L = z_i(x\leq 0)$ and $z_i^R = z_i(x\geq 0)$ the components of the eigenvectors to the left and right of $x = 0$. The boundary conditions can be derived with minimal algebra
\begin{equation}
\label{eq:D2BCs}
\begin{cases}
& \partial_x q|_{x= \pm L_0/2} = 0 \Leftrightarrow \displaystyle  q_{\rm off} z_1^{L/R} |_{x= \pm L_0/2} =  z_2^{L/R} |_{x= \pm L_0/2} \\
& \mathrm{continuity \, of} \, p \Leftrightarrow \displaystyle \frac{ q_{\rm on} z_1^{L/R}  + z_2^{L/R}}{q_{\rm off} + q_{\rm on}}\bigg|_{x= \pm L_0/2} =  z_m^{L/R} |_{x= \pm L_0/2} \\
& \mathrm{continuity \, of}  \,\partial_x p \Leftrightarrow \displaystyle  \partial_x\frac{ q_{\rm on} z_1^{L/R}  + z_2^{L/R}}{q_{\rm off} + q_{\rm on}}\bigg|= \partial_x z_m^{L/R} |_{x= \pm L_0/2} 
\end{cases}
\end{equation}

\paragraph{Starting from the left.} 
We now define the Laplace transforms
\begin{equation}
    \hat{z}^{L/R}_i(x,s) = \int_0^{\infty} e^{-st} \left( \tilde{z}^{L/R}_i(x,t) - \tilde{z}^{L/R}_i(x,t=0) \right) dt.
\end{equation}
Note that here the Laplace transforms are taken with respect to the base value. The general solutions of the partial differential functions aforementioned are  
\begin{equation}
    \hat{z}^{L/R}_i = A^{L/R}_i e^{q_i x} + B^{L/R}_i e^{-q_i x}
\end{equation}
where $q_1 = q_m = q = \sqrt{\frac{s}{D}}$ and $q_2 = \tilde{q} = \sqrt{q^2 + \frac{q_{\rm off}+ q_{\rm on}}{D}}$. And the integration constants $A_i, B_i$ are determined by the boundary conditions Eq.~\eqref{eq:D2BCs}. We recall that the contributions to the noise from this first situation are 
\begin{equation}
 \langle    \Delta N^2(t) \rangle^{(1)} =  8N\int_{L_0/2}^{L/2} \left[p(x,t)+q(x,t)\right] dx   + 2N\int_{-L_0/2}^{L_0/2} p(x,t) dx
\end{equation}
such that
\begin{equation}
    \begin{split}
         \partial_t \langle    \Delta N^2(t) \rangle^{(1)} &=  8N\int_{L_0/2}^{L/2} \partial_t z^R_1 dx   + 2N\int_{-L_0/2}^{L_0/2} \partial_t z_m dx  \\
         &=  8N\int_{L_0/2}^{L/2} D\partial_{xx} z_1^R dx   + 2N\int_{-L_0/2}^{L_0/2} D \partial_{xx} z_m dx    \\
         &= - 8N\, D\partial_x\left[z_1\right]_{x=L_0/2} + 2N\, D\partial_x\left[z_m\right]_{x=L_0/2} \\
         & \,\,\,\,\,\,\,\,\,\,\,\,\,\,\, - 2N\, D\partial_x\left[z_m\right]_{x=0} 
    \end{split}
\end{equation}
We thus obtain in Laplace space
\begin{equation}
\begin{split}
    \langle    \hat{\Delta N^2}(s) \rangle^{(1)} &= 8 \frac{N}{q} \left( B_1^R e^{-q L_0/2} - A_1^R e^{qL_0/2} \right) + \\
         & \,\,\, + 2 \frac{N}{q}  \left( A_m e^{-qL_0/2} - B_m e^{qL_0/2} - A_m e^{qL_0/2} + B_m e^{qL_0/2} \right)
\end{split}
\end{equation}
Starting from the left side, $z_1^L(t=0) = \displaystyle \frac{1}{L}\frac{1}{1 - \frac{L_0}{L}\left( \frac{1}{1+r}\right)}$ where $r = q_{\rm on}/q_{\rm off}$ and otherwise intial conditions are zero.

\paragraph{Starting from the channel.}
We can solve in a similar way for particles starting from the center of the domain. In that case the problem is symmetric with respect to the center $x = 0$ and can be simplified accordingly. Using similar notations we obtain
\begin{equation}
         \partial_t \langle    \Delta N^2(t) \rangle^{(2)} =  2 N\int_{L_0/2}^{L/2} \partial_t z^R_1 dx  
\end{equation}
and in Laplace space
\begin{equation}
    \langle    \hat{\Delta N^2}(s) \rangle^{(2)} =  2 \frac{N}{q} \left( B_1^R e^{-q L_0/2} - A_1^R e^{qL_0/2} \right) 
\end{equation}
Starting from the channel the base conditions are zero everywhere except for $z_m(t=0) = \displaystyle \frac{r}{1+r} \frac{1}{L}\frac{1}{1 - \frac{L_0}{L}\left( \frac{1}{1+r}\right)}$. 

The total fluctuations sum up to 
\begin{equation}
\langle  \hat{\Delta N^2}(t) \rangle= \langle  \hat{\Delta N^2}(t) \rangle^{(1)} + \langle  \hat{\Delta N^2}(t) \rangle^{(2)} 
\end{equation}

\paragraph{Correlations of the number of particles within the channel}
The spectrum of the correlation function of the number of particles within the channel can be simply inferred starting from 
\begin{equation}
\langle N_c(t)N_c(0) \rangle = 2\int_0^{L_0/2} z_m(x,t) dx 
\end{equation}
and going to Laplace space we simply obtain
\begin{equation}
\mathcal{L}\left[ \langle N_c(t) N_c(0) \right] (s) = -2 \frac{D}{s}  \partial_x z_m\big|_{x = L_0/2}
\end{equation}

\subsubsection*{D. 3. Solutions in Laplace space}

\begin{figure}[h]
\includegraphics[width = 0.90\columnwidth]{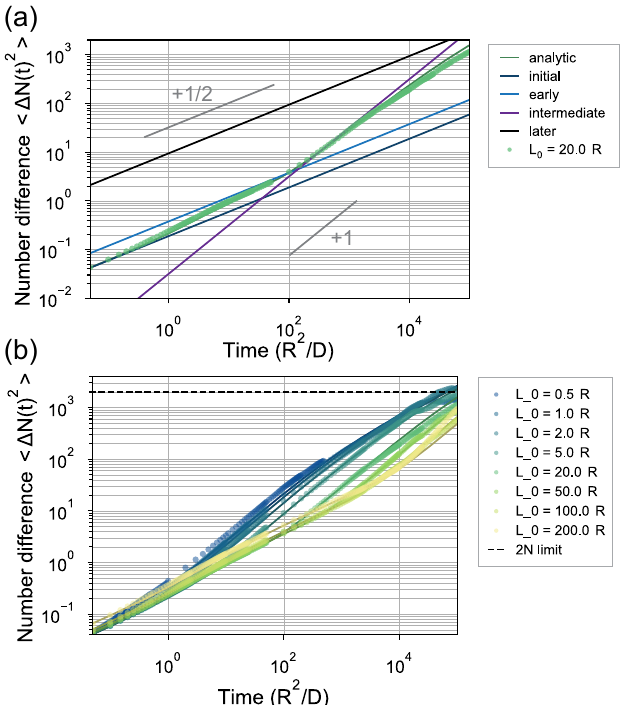}
\caption{\label{fig:appendix2} \textbf{The rates model reproduces the regimes for the number difference fluctuations observed in a long 3D square pore}. Here $R = 0.2 L_m$ but similar results were found also for much small pores. (a) Different regimes with time, as illustrated in Fig.~\ref{fig:ChannelIntro}-a, for a specific channel geometry. Initial corresponds to Eq.~\ref{eq:limit3DThickDNinitial}, early to Eq.~\ref{eq:limit3DThickDNearly}, intermediate to Eq.~\ref{eq:limit3DThickDNintermediate} and later to Eq.~\ref{eq:limit3DThickDNlate}. The later regime is not so much observed as it corresponds to a time point where fluctuations reach the saturation limit $2N$.  (b) Number difference fluctuations with time for several values of $L_0/R$ as indicated with the various colors. Dots correspond to data from BD and lines to the analytic solutions Eq.~\eqref{eq:ChannelLDN}.  Numerical parameters correspond to that of Fig.~\ref{fig:fig1}.}
\end{figure}

Here we report only the results when formally $L \rightarrow \infty$, corresponding to times $t \ll L^2/D$. This imposes $A_{1/2}^{R/L} = 0$ and therefore simplifies greatly the problem leaving only 6 integration constants to be found. 
We write $r = q_{\rm on}/q_{\rm off}$ and $r(s) = q/\tilde{q}$ such that 
\begin{equation}
\begin{split}
\langle & \hat{\Delta N^2}(s) \rangle=  4 N \frac{\sqrt{D}}{s^{3/2} L} r (1+r) (1 + e^{L_0q} ) / \\
&\,\,\, \bigg( \left[ 1+r - L_0/L \right] \times \left[ r(q) -1 + e^{L_0 q}(r(q) + 1 + 2r) \right] \bigg)
\end{split}
\label{eq:ChannelLDN}
\end{equation}
Similarly 
\begin{equation}
\begin{split}
\langle  & \hat{N_c^2}(s) \rangle=  4 N \frac{\sqrt{D}}{s^{3/2} L} r (1+r) (- 1 + e^{L_0q} ) / \\
&\,\,\, \bigg( \left[ 1+r - L_0/L \right] \times \left[ 1- r(q) + e^{L_0 q}(r(q) + 1 + 2r) \right] \bigg).
\end{split}
\label{eq:ChannelLNc}
\end{equation}
These solutions recover in particular the expected limits for infinitely thin pores $L_0 = 0$ and infinitely long channels $q_{\rm off} = 0$. Agreement of these analytic results with BD simulations is reported in Figs.~\ref{fig:appendix2} and~\ref{fig:appendix3}.

\subsubsection*{D. 4. Limit regimes}

\paragraph{Narrow pores}
We start by investigating the limit regimes in narrow pores ($r \rightarrow 0$). 
We obtain
\begin{equation}
\begin{split}
& \langle  \hat{\Delta N^2}(s) \rangle=  4 N \frac{\sqrt{D}}{s^{3/2} L} r  (1 + e^{L_0q} ) / \\
&\,\,\, \bigg( \left[ 1- L_0/L \right] \times \left[ r(q) -1 + e^{L_0 q}(r(q) + 1) \right] \bigg)
\end{split}
\end{equation}
that is easily amenable to early/intermediate/late times investigation. 
We start by short times, corresponding to $s, q \rightarrow \infty$. We find
\begin{equation}
\begin{split}
\langle  \hat{\Delta N^2}(s) \rangle= & 2 N \frac{\sqrt{D}}{s^{3/2} } \frac{r}{L- L_0} 
\end{split}
\end{equation}
such that the real time evolutions scales as 
\begin{equation}
\langle \Delta N^2(t) \rangle \underset{t \ll L_0^2/D}{=} 4\frac{N}{L-L_0} \frac{\pi R^2}{4L_m^2}\sqrt{\frac{D t}{\pi}}
\label{eq:limit3DThickDNinitial}
\end{equation}
and similarly for the number of particles inside the channel
\begin{equation}
\langle N_c^2(t) \rangle \underset{t \ll L_0^2/D}{=} 4\frac{N}{L-L_0} \frac{\pi R^2}{4L_m^2}\sqrt{\frac{D t}{\pi}}.
\label{eq:limit3DThickNcinitial}
\end{equation}

At intermediate times, corresponding to $L_0 q \rightarrow 0$ but $q/q_{\rm off} \rightarrow \infty$ we obtain
\begin{equation}
\langle  \hat{\Delta N^2}(s) \rangle=  4 N \frac{\sqrt{D}}{s^{3/2} L} \frac{r}{L- L_0} 
\end{equation}
yielding 
\begin{equation}
\langle \Delta N^2(t) \rangle \underset{t \gtrsim L_0^2/D}{=} 8\frac{N}{L-L_0} \frac{\pi R^2}{4L_m^2}\sqrt{\frac{D t}{\pi}}
\label{eq:limit3DThickDNearly}
\end{equation}
and similarly for the number of particles inside the channel
\begin{equation}
\langle N_c^2(t) \rangle \underset{t \gtrsim L_0^2/D}{=} 8\frac{N}{L-L_0} \frac{\pi R^2}{4L_m^2}\sqrt{\frac{D t}{\pi}}.
\label{eq:limit3DThickNcearly}
\end{equation}

Going a bit further in time, we can investigate the infinite time limit for the number of particles, $q \rightarrow 0$. We obtain, as expected
\begin{equation}
\langle  \hat{N_c^2}(s) \rangle=  2 N \frac{L_0}{s L}  \frac{1}{  1 - L_0/L }
\end{equation}
yielding typically 
\begin{equation}
\langle N_c^2(t) \rangle \underset{t \gg L_0^2/D}{=} 2N_0
\label{eq:limit3DThickNcLimit}
\end{equation}
where $N_0$ is the mean number of particles inside the channel. 

\begin{figure}[h]
\includegraphics[width = 0.90\columnwidth]{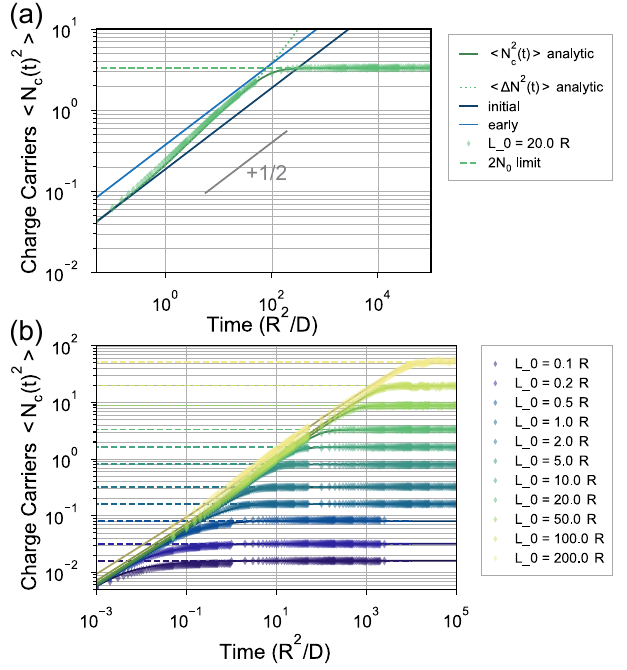}
\caption{\label{fig:appendix3} \textbf{The rates model reproduces the regimes for the number of particles within the pore fluctuations observed in a long 3D square pore}. Here $R = 0.2 L_m$ but similar results were found also for much small pores. (a) Different regimes with time, as illustrated in Fig.~\ref{fig:ChannelIntro}-a, for a specific channel geometry. Initial corresponds to Eq.~\ref{eq:limit3DThickNcinitial}, early to Eq.~\ref{eq:limit3DThickNcearly}, and saturation to the $2 N_0$ limit of Eq.~\ref{eq:limit3DThickNcLimit}. Analytic solutions for $\langle N_c^2 (t) \rangle$ and $\langle \Delta N^2(t) \rangle$ overlap at short times. (b) Number difference fluctuations with time for several values of $L_0/R$ as indicated with the various colors. Dots correspond to data from BD and lines to the analytic solutions Eq.~\eqref{eq:ChannelLNc}.  Numerical parameters correspond to that of Fig.~\ref{fig:fig1}. }
\end{figure}

Coming back to the number difference at these intermediate times, taking now $q/q_{\rm off} \rightarrow 0$ (but not too small...) 
\begin{equation}
\langle  \hat{\Delta N^2}(s) \rangle=  4 N \frac{\sqrt{D}}{s^{3/2} } \frac{r}{L-L_0} \frac{\sqrt{q_{\rm off}/D}}{q} 
\end{equation}
such that we recover the intermediate, diffusive regime as
\begin{equation}
\langle \Delta N^2(t) \rangle \underset{t \gtrsim q_{\rm off}^{-1}}{=} 4 \frac{N}{L-L_0} \frac{\pi R^2}{4L_m^2} \sqrt{Dq_{\rm off}} t
\label{eq:limit3DThickDNintermediate}
\end{equation}
and finally at very long times ($q \rightarrow 0$)
\begin{equation}
\langle  \hat{\Delta N^2}(s) \rangle= 4 N \frac{\sqrt{D}}{s^{3/2} L} \frac{1}{1-L_0/L}
\end{equation}
yielding the natural 
\begin{equation}
\langle \Delta N^2(t) \rangle \underset{t \gg q_{\rm off}^{-1}}{=} 8 \frac{N}{L-L_0}  \sqrt{\frac{D t}{\pi} }.
\label{eq:limit3DThickDNlate}
\end{equation}
The emergence of all these limit regimes is shown in Figs.~\ref{fig:appendix2}-b and~\ref{fig:appendix3}-b.

\subsubsection*{D. 5. Comparison of the results with another method}

\begin{figure}[h!]
\includegraphics[width = 0.65\columnwidth]{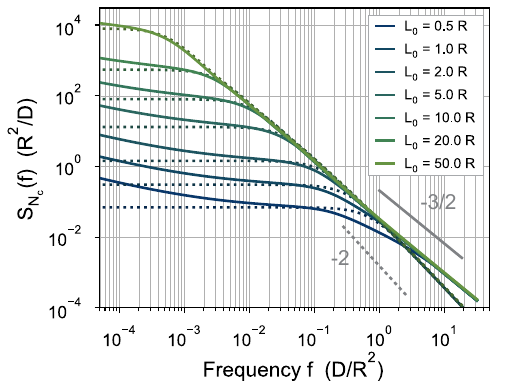}
\caption{\label{fig:figBeru} \textbf{Comparison of different methods}. Fluctuation spectrum for the number of particles within the pore calculated using Eq.~\eqref{eq:SNcS3D} (full line) and with Eq.~(12) of Ref.~\onlinecite{bezrukov2000particle} (dotted line). Here we took $R = 0.2 \, L_m$ and $L = 500 \, R$. To compare with Eq.~(12) of Ref.~\onlinecite{bezrukov2000particle} we naturally used $D_b = D$ in their notations. Note that Eq.~(12) had to be divided by a factor 2 to show some degree of matching with our result Eq.~\eqref{eq:SNcS3D}.}
\end{figure}

In this paragraph we compare the results of Eq.~\eqref{eq:SNcS3D} with the results obtained with Eq.~(12) of Ref.~\onlinecite{bezrukov2000particle}. Fig.~\ref{fig:figBeru} shows the two solutions plotted for representative parameters. Disagreement between the two solutions is visible at low frequencies, at the frequency turning point, and at high frequencies were the decay exponent is not the same. Similar disagreement is found for other pore parameters. 

\end{document}